\documentclass[aps,prl,twocolumn,superscriptaddress,secnumarabic]{revtex4-2}

\usepackage{graphicx}
\usepackage[utf8]{inputenc}
\usepackage{float}
\usepackage{comment}

\usepackage{xcolor}
\usepackage{amsmath}

\begin{document}

\title{On the origin of circular dichroism in angular resolved photoemission from graphene, graphite, and WSe$_2$ family of materials}

\author{L. Plucinski}
\affiliation{Peter Gr{\"u}nberg Institut (PGI-6), Forschungszentrum J{\"u}lich GmbH,
52428 J{\"u}lich, Germany}


\date{\today}

\begin{abstract}
Circular dichroism in angle-resolved photoemission (CD-ARPES) is one of the promising techniques for obtaining experimental insight into topological properties of novel materials, in particular to the orbital angular momentum (OAM) in dispersive bands, which might be related, albeit certainly in a non-trivial way, to the momentum resolved Berry curvature of the bands. Therefore, it is important to understand how non-vanishing CD-ARPES signal arises in graphene, a material where Dirac bands are made from C $|2p_z\rangle$ orbitals that carry zero OAM, spin-orbit-coupling (SOC) can be neglected, and Berry curvature effectively vanishes. Dubs et al., Phys. Rev. B 32, 8389 (1985) have demonstrated non-vanishing cricular dichroism in angular distribution (CDAD) from an oriented $p_z$ orbital, and this process can be responsible for the experimentally observed CD-ARPES in graphene. In this paper, we derive the CD-ARPES from $p_z$ orbitals by elementary means, using only simple algebraic formulas and tabulated numerical values, and show that it leads to significant CD-ARPES signal over the entire vacuum ultraviolet and soft x-ray energy range, with an exception of the photon energy region near $h\nu \approx 40$ eV. We also demonstrate that another process, emerging from the finite electron inelastic mean free path, also leads to CD-ARPES of the potentially similar order of magnitude, as previously discussed by Moser, J. Electron Spectrosc. Relat. Phenom. 214, 29 (2017). We present calculated CDAD maps for selected orbitals and briefly discuss the consequences of the findings for CD-ARPES, focusing on graphene, graphite and WSe$_2$.
\end{abstract}

\maketitle

\section{Motivation, historical perspective, and the formalism}

Circular dichroism in angle-resolved photoemission (CD-ARPES), together with spin-polarized ARPES (SARPES) may allow an experimental access to the momentum dependent Berry curvature of the Bloch bands, a critical quantity which enters the Kubo formula for quantum transport. According to the tight binding toy models \cite{Nagaosa2010RMP,Xiao2010RMP}, a non-vanishing Berry curvature appears in bands that exhibit both non-vanishing spin-polarization and orbital angular momentum (OAM), for which sizeable spin-orbit coupling is typically necessary. Therefore, in order to make ARPES useful in deriving properties of Berry curvature, it is important to understand the origin of CD-ARPES from graphene, a material where SOC can be neglected because of its energy scale being much smaller than the experimental resolution of CD-ARPES. Experimentally, CD-ARPES from graphene has been reported \cite{Liu2011,Gierz2012} and recently, numerical calculations based on the one-step-model of photoemission \cite{Krueger2022} have also been performed. In the context of atomic and molecular physics, CD-ARPES has been typically called circular dichroism in angular distribution (CDAD), here we will use these acronyms interchangeably.

Original work of atomic photoionization dates back to Hans Bethe \cite{Bethe1933}, with the compact formula provided by Cooper and Zare \cite{Cooper1968}, and early work reviewed by Fano and Cooper \cite{Fano1968}. Dubs et al. derived another formula for CDAD \cite{Dubs1985PRL} and subsequently, in 1985, \cite{Dubs1985} they have theoretically predicted non-vanishing CDAD signal from an oriented $p_z$ orbital. This kind of CDAD process, which does not require SOC, has been confirmed experimentally by Westphal et al. \cite{Westphal1989} by measuring oriented CO molecules on Pd(111), and further discussed in detail by Sch\"onhense \cite{Schonhense1990}, also in the context of dispersive valence bands. There also exists previous work on CD-ARPES in core level spectroscopy \cite{Daimon1993,Kaduwela1995}, and early theoretical work on CDAD from chiral molecules \cite{Cherepkov1982}. The literature on atomic and molecular photoionization is very extensive, and the above account is certainly selective.

The photoionization derivations have typically used hydrogenic atomic orbitals as initial states, and Coulomb scattering state for the final state. Based on the previous work, Moser \cite{Moser2023} has recently revived the scattering final state formalism focusing on CD-ARPES from a crystalline solid and here we follow this formalism. It is taken as an assumption, that one can use scattering states for all sites, and combine them as a linear combination. The detailed relation of this model to the three-step photoemission model, where bound states of the crystal are used as final states, needs to be established, however, it will not be discussed here. Some of these issues have been previously discussed in Ref. \cite{Schonhense1990}. As discussed in H\"ufner's textbook \cite{Huefner2003} in the chapter on photoelectron diffraction, eventually, both formalisms should converge to the same result, if multiple scattering is considered. As a first attempt towards this, very recently, Kern et al. \cite{Kern2023} have included nearest neighbor scattering in the graphene photoemission model. The numerically-heavy one step model of photoemission allows proper treatment of these issues, however, in this manuscript the focus is on providing elementary means to understand the origin of the CD-ARPES effects. Dichroic magneto-optical processes have been described for example in Refs. \cite{Ebert1996,Henk1996}.

Angular part $Y_l^m$ of an atomic orbital is classified by $l$ and $m$ quantum numbers. Angularly integrated net CD signal from $m=0$ orbitals disappears, while the net CD signal from oriented $m\neq 0$ orbitals is in general non-zero. This is relevant to absorption or total-electron-yield measurements such as x-ray magnetic circular dichroism (XMCD), however, for CD-ARPES the angular distribution is critical. Here, we will not consider final state (multiple) scattering, and we will assume that the angle-resolved photoemission signal is composed of dispersions in emitted electrons $B(\mathbf k_f)$, modulated by an envelope (atomic form factor) of the angular distribution from an atomic orbital $D(\mathbf k_f)$

\begin{equation}
\begin{aligned}
I(k_f) = D(\mathbf k_f) \cdot B(\mathbf k_f)
\end{aligned}
\end{equation}

At least for 2D materials, $B(\mathbf k_f)$ can be faithfully connected to the initial band dispersion $B(\mathbf k_i)$ through the parallel momentum conservation $\mathbf k_{i||} = \mathbf k_{f||}$. Multiple scattering, if taken into account, would modify the atomic envelope function $D(\mathbf k_f)$. The easiest application of the model is then for the bands that exhibit dominant single-orbital character. One example is the Dirac cone in graphene composed from C $|2p_z\rangle$, with only $\approx 2\%$ admixture of C $|3d\rangle$ states. Another example is a monolayer of WSe$_2$, where orbitals at $K$ and $K'$ are made from W $5d$ orbitals of $Y_2^{\pm 2}$ angular character, while those at $\Gamma$ primarily from the W $5d$ $Y_2^0$ orbitals. Therefore, in this paper we focus on function $D(\mathbf k_f)$ to obtain CDAD from oriented orbitals.

The atomic scattering wave function is written as a partial wave expansion \cite{Moser2023,Goldberg1981}

\begin{equation}
\begin{aligned}
& \psi_f(\mathbf r,\mathbf{k}_f) = \\
& 4\pi \sum_{l=0}^\infty \sum_{m=-1}^1 i^l e^{i\sigma_l(E_{kin})} R_{E_{kin},l}(r)~Y_l^{m*}(\mathbf {\hat r}) ~ Y_l^m(\mathbf{\hat k}_f)
\label{eq:partial_wave}
\end{aligned}
\end{equation}

where $^*$ denotes complex conjugation, $\mathbf{\hat r} = \mathbf{r}/|r|$, and $\mathbf{\hat k}_f = \mathbf{k}_f/|k_f|$. This formula is similar to the partial wave expansion for the plane wave $e^{i\mathbf k \cdot \mathbf r}$, however, contains Coulomb phase shifts $\sigma_l(E_{kin})$, which depend on $l$ and on kinetic energy \cite{Goldberg1981}, and the radial part of the wave function is not a Bessel function.


We assume that the initial state Bloch wave function can be written as a linear combination of atomic orbitals

\begin{equation}
\begin{aligned}
\psi_i(\mathbf r,\mathbf k_i) = \sum_{n,l,m,\mathbf R} c_{nlm, \mathbf R}(\mathbf k_i)~\phi_{nlm}(\mathbf r - \mathbf R)
\label{eq:Bloch}
\end{aligned}
\end{equation}

The atomic inital wave function can be factorized into radial and angular parts

\begin{equation}
\begin{aligned}
\phi_{nlm} = R_{nl} ~ Y_l^m
\end{aligned}
\end{equation}

The matrix element for photoemission in the length form reads

\begin{equation}
\begin{aligned}
M_{fi}(\mathbf k_f) \propto \langle \psi_f(\mathbf k_f) | \boldsymbol{\varepsilon} \cdot \mathbf{r} | \psi_i \rangle
\end{aligned}
\end{equation}

where $\boldsymbol \varepsilon$ is the light polarization vector and $\mathbf r$ is the position operator.

Since the matrix element is a space integral, and the initial state wave function in Eq. \ref{eq:Bloch} is written as a sum over all initial state orbitals, the matrix element is also a sum of matrix elements for each initial orbital separately. Therefore, even though in the following we will only consider a single initial orbital $\phi_{nlm}$, the formalism can be expanded for the complete tight binding wave function.

\begin{table}
\begin{tabular}{| l  c | }
\hline \hline
$C_+$ light & \\
\hline
$C_{l\rightarrow l+1}^{m\rightarrow m+1} = \sqrt{ \frac{(l+m+2)(l+m+1)}{2(2l+3)(2l+1)} }$ &  \\
$C_{l\rightarrow l-1}^{m\rightarrow m+1} = -\sqrt{ \frac{(l-m)(l-m-1)}{2(2l-1)(2l+1)} }$ &  \\
\hline \hline
$C_-$ light & \\
\hline
$C_{l\rightarrow l+1}^{m\rightarrow m-1} = \sqrt{ \frac{(l-m+2)(l-m+1)}{2(2l+3)(2l+1)} }$ &  \\
$C_{l\rightarrow l-1}^{m\rightarrow m-1} = -\sqrt{ \frac{(l+m)(l+m-1)}{2(2l-1)(2l+1)} }$ &  \\
\hline \hline
\end{tabular}
\caption{\label{table:Stohr} Angular matrix elements for $C_\pm$ light following St\"ohr and Siegmann \cite{Stohr-Siegmann}, page 381.}
\end{table}

The length-form matrix element can be factorized to separately calculate the radial integral and the angular integral using spherical coordinate system. We know that angular integrals

\begin{equation}
\begin{aligned}
C_{l\rightarrow l\pm 1}^{m\rightarrow m\pm 1} =  \langle Y_{l\pm 1,m\pm 1} | C_\pm | Y_{lm} \rangle
\end{aligned}
\end{equation}

only allow for $l\pm 1$ transitions leading to selection rules and their values are simple algebraic formulas listed for example in the St\"ohr-Siegmann textbook \cite{Stohr-Siegmann} and we list the ones for $C_\pm$ light in Table \ref{table:Stohr}. This also means that radial integrals for only $l\pm 1$ need to be taken into account. One can also see that the $Y_l^m(\mathbf{\hat k}_f)$ in Eq. \ref{eq:partial_wave}, that does not depend on the position, will determine the angular distribution of the $l\pm 1$ channels of the matrix element.

\begin{figure}
 \centering
     \includegraphics[width=8cm]{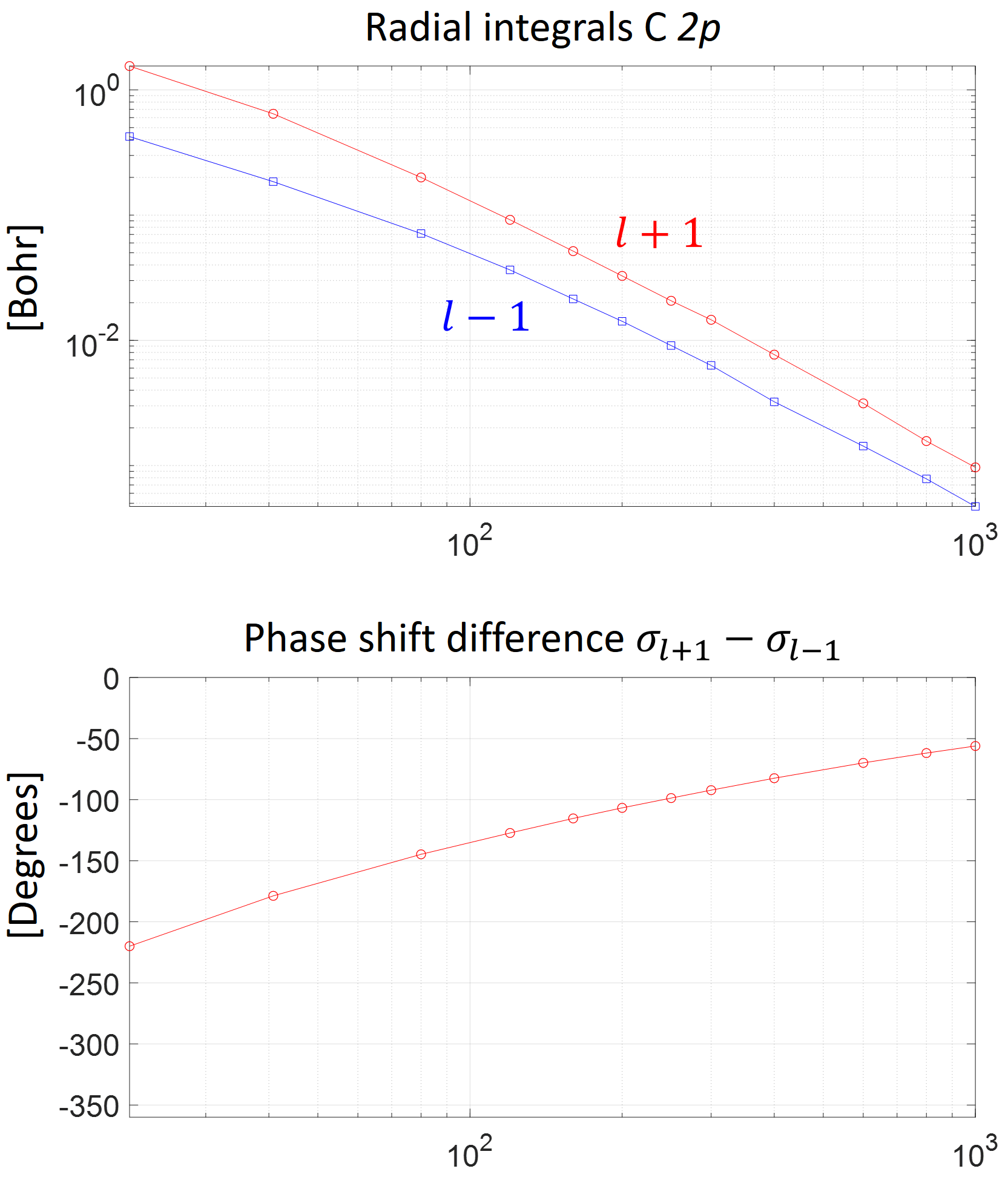}
     \caption{Radial integrals and Coulomb phase shift differences for C $|2p\rangle$ for energies between 21.2 and 1000 eV. Plotted data is taken from Goldberg et al. \cite{Goldberg1981}.}
     \label{fig:Goldberg}
\end{figure}

Radial integrals take the form \cite{Goldberg1981}

\begin{equation}
\begin{aligned}
R_{l\rightarrow l\pm 1}(E_{kin}) = \int_{0}^{\infty} r^2 R_{nl} r R_{E_{kin},l\pm 1} dr
\label{eq:radial}
\end{aligned}
\end{equation}

and have been tabulated, together with the related Coulomb phase shifts, for selected photon energies by Goldberg, Fadley, and Kono \cite{Goldberg1981}. In Fig. \ref{fig:Goldberg} we plot them, as well as phase shift differences, for C $|2p\rangle$ for the energy range between 21.2 and 1000 eV. The factor $r^2$ appears in Eq. \ref{eq:radial} because the space integration is performed in the spherical basis $(x,y,z) \rightarrow (r,\theta,\varphi)$.

Therefore, having the radial and angular integrals, and the Coulomb phase shifts  is sufficient to calculate the angular photoemission intensity from any orbital, which is the square of the amplitude of the matrix element. This reduces the calculation to elementary algebra, with all components and factors either given explicitly as a function of $l$ and $m$ (angular part, Table \ref{table:Stohr}) or tabulated e.g. in Goldberg et al. \cite{Goldberg1981} (phase shifts, radial integrals). The $l+1$ channel is always available, the $l-1$ channel might not be available, for example $C_+$ light would emit from $Y_1^0$ into $Y_0^1$, and such final state orbital does not exist. It is therefore evident which final states are available.

For example the matrix element for emission from $Y_l^m$ orbital into the $l+1$ channel with $C_+$ light reads

\begin{equation}
\begin{aligned}
& M_{l\rightarrow l+1,C_+} (\mathbf{k}_f) \propto \\
& i^{l+1} e^{i\sigma_{l+1}(E_{kin})} \cdot R_{l\rightarrow l+1}(k_f) \cdot C_{l\rightarrow l+1}^{m\rightarrow m+1} \cdot Y_{l+1}^{m+1}(\mathbf{\hat k}_f) \\
\end{aligned}
\end{equation}

This matrix element depends on the kinetic energy trough the $E(k_f)$ dependence of the radial integral and phase shift. Note, that the $Y_{l+1}^{m+1}(\mathbf{\hat k}_f)$ does not depend on $E_{kin}$, and it determines the angular distribution of the emitted electrons for that channel, as mentioned earlier. For a particular $\mathbf k_f$, $Y_{l+1}^{m+1}(\mathbf{\hat k}_f)$ is a constant factor, and, for each $l\pm 1$ channel, it can be taken outside of the matrix element space integral .

In the following sections we will compute CDAD from selected orbitals using the formalism described above. We will also compute another type of CDAD which originates from the inelastic mean free part (IMFP) of the orbitals submerged in the electron gas of other electrons near the surface of the solid \cite{Moser2017}.

Matrix elements are complex numbers. If the state is made out of a combination of various orbitals, then for each light polarization one needs to coherently sum matrix element contributions from all these orbitals into all the available $l\pm 1$ channels, and, only after that, square that sum $M$ as $M^* M$ for get the intensity. If orbitals on different sites are considered, then in addition the phase differences along the direction of the electron emission must be considered \cite{Heider2023}. This is meaningful, because in the solid, the wave function phases on different sites are well defined.

\begin{figure*}
 \centering
     \includegraphics[width=16cm]{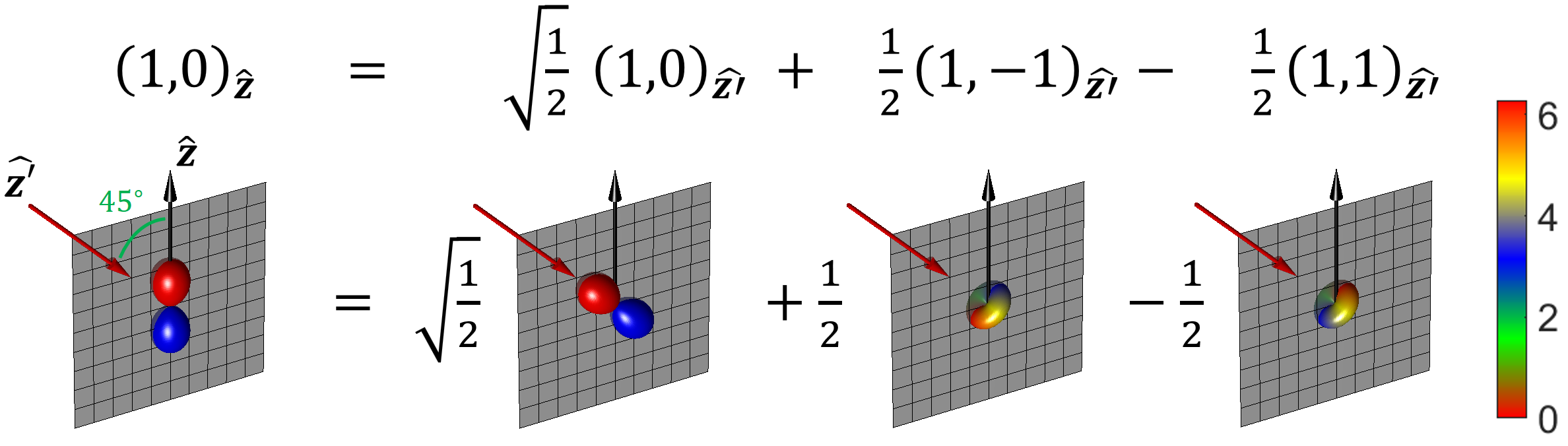}
     \caption{Graphical visualization of the coordinate system rotation by 45$^\circ$ for the $p_z$ orbital. The orbital axis (black arrow) is depicted as $\mathbf{\hat z}$, and the light incidence direction (red arrow) as $\mathbf{\hat z'}$. Colorbar on the right side indicates the orbital phase in radians.}
     \label{fig:pz_rotation}
\end{figure*}

\begin{figure}
 \centering
     \includegraphics[width=8cm]{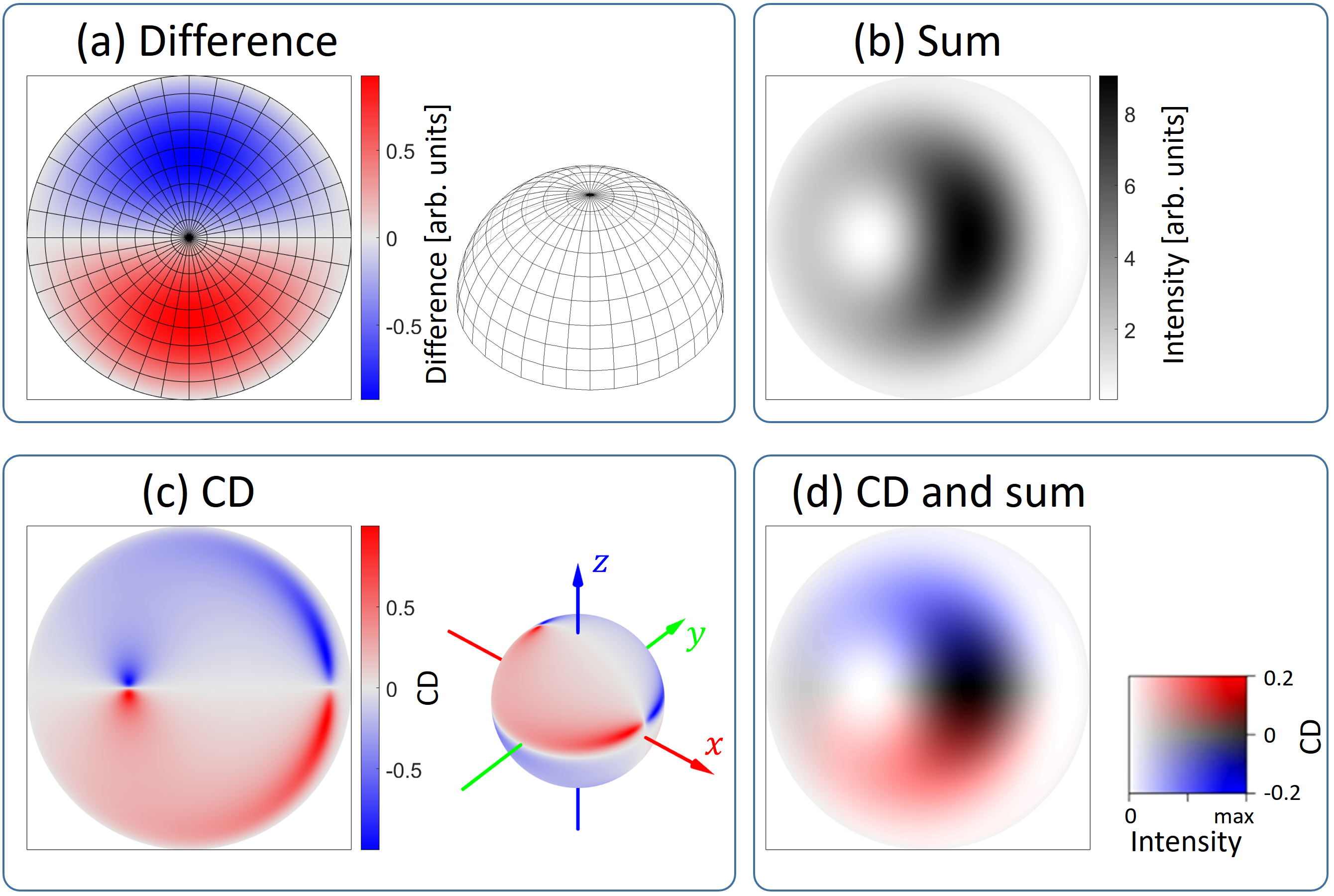}
     \caption{Angular depencence of photoemission from C$|2p\rangle$ orbital with light incidence at $45^\circ$ using the numerical parameters from Goldberg et al. \cite{Goldberg1981} for $h\nu = 80$ eV. (a) Difference between angular distributions with $C_+$ and $C_-$ light. The image also includes the $10^\circ$ grid both in polar and azimuthal orientation, and the 3D impression of the grid, visualizing the way the projection is performed. (b) Sum of the angular distributions taken with $C_+$ and $C_-$ light. (c) Circular dichroism, difference divided by the sum at each angle. The panel also includes the 3D visualization of the CD angular distribution. (d) Combination of (b) and (c) using a 2D colormap shown at the right side of the panel, with the color scale saturated to $\pm 0.2$. Images show projection of the half sphere centered on the $\mathbf{\hat z}$ axis, with axes indicating emission angles in degrees. Units of color scales in (a) and (b) are arbitrary, but the same in both cases, thus allowing comparison.}
     \label{fig:CDAD_2pz_80eV}
\end{figure}

\begin{figure}
 \centering
     \includegraphics[width=8cm]{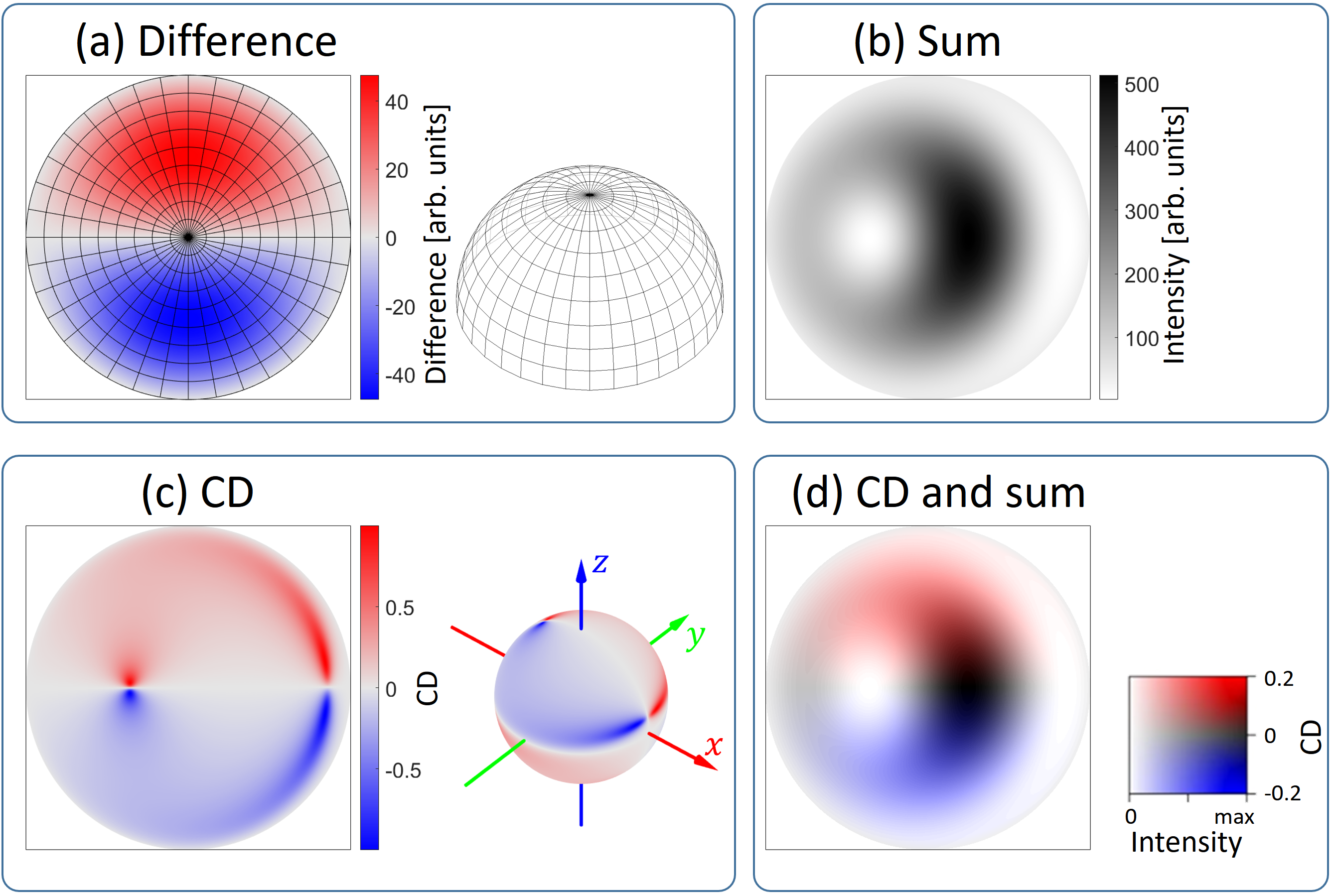}
     \caption{Same as Fig. \ref{fig:CDAD_2pz_80eV} but for $h\nu = 21.2$ eV.}
     \label{fig:CDAD_2pz_21eV}
\end{figure}

\begin{figure}
 \centering
     \includegraphics[width=8cm]{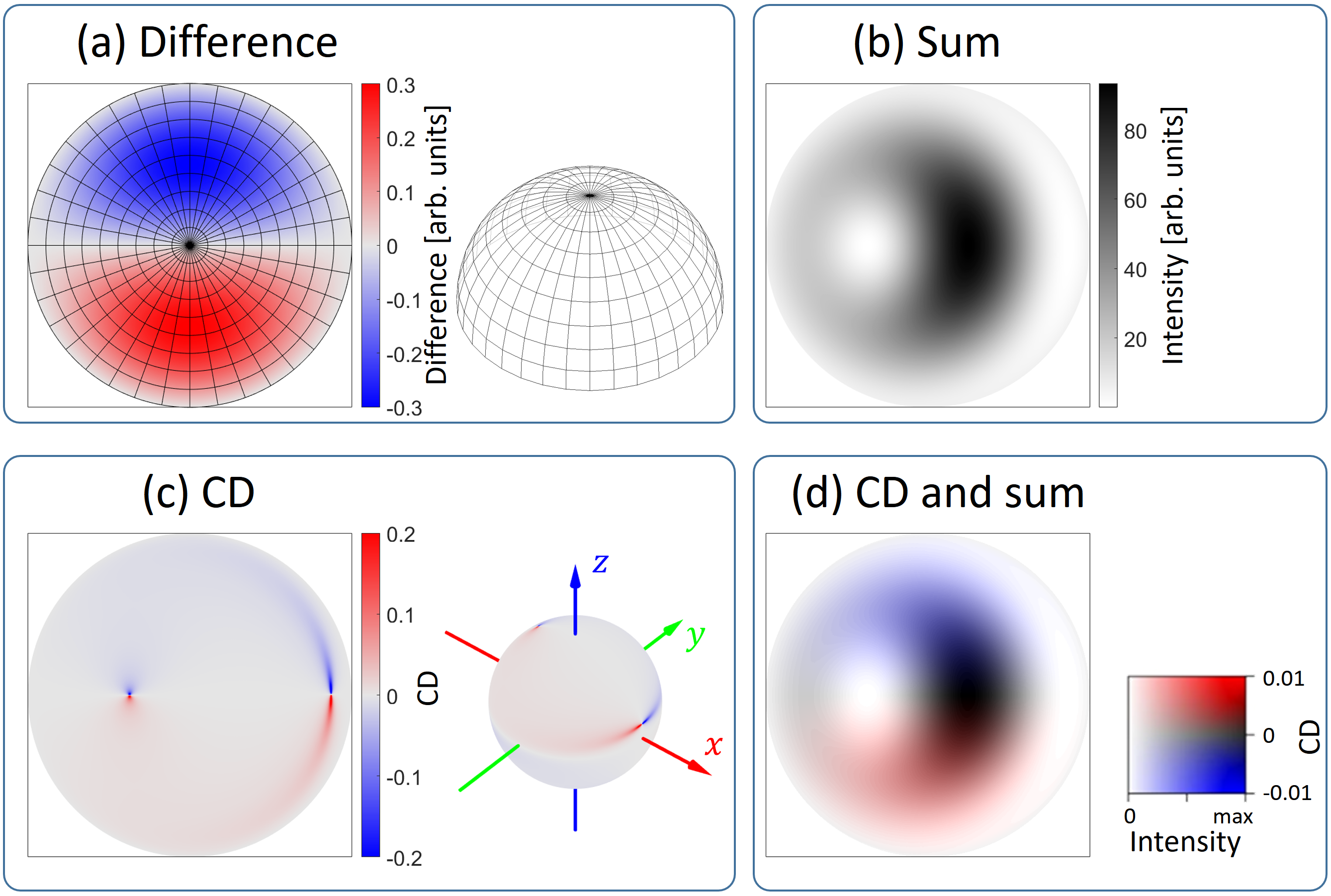}
     \caption{Same as Fig. \ref{fig:CDAD_2pz_80eV} but for $h\nu = 40.8$ eV. Note the color scale normalization in panels (c) and (d).}
     \label{fig:CDAD_2pz_40eV}
\end{figure}

\begin{figure}
 \centering
     \includegraphics[width=8cm]{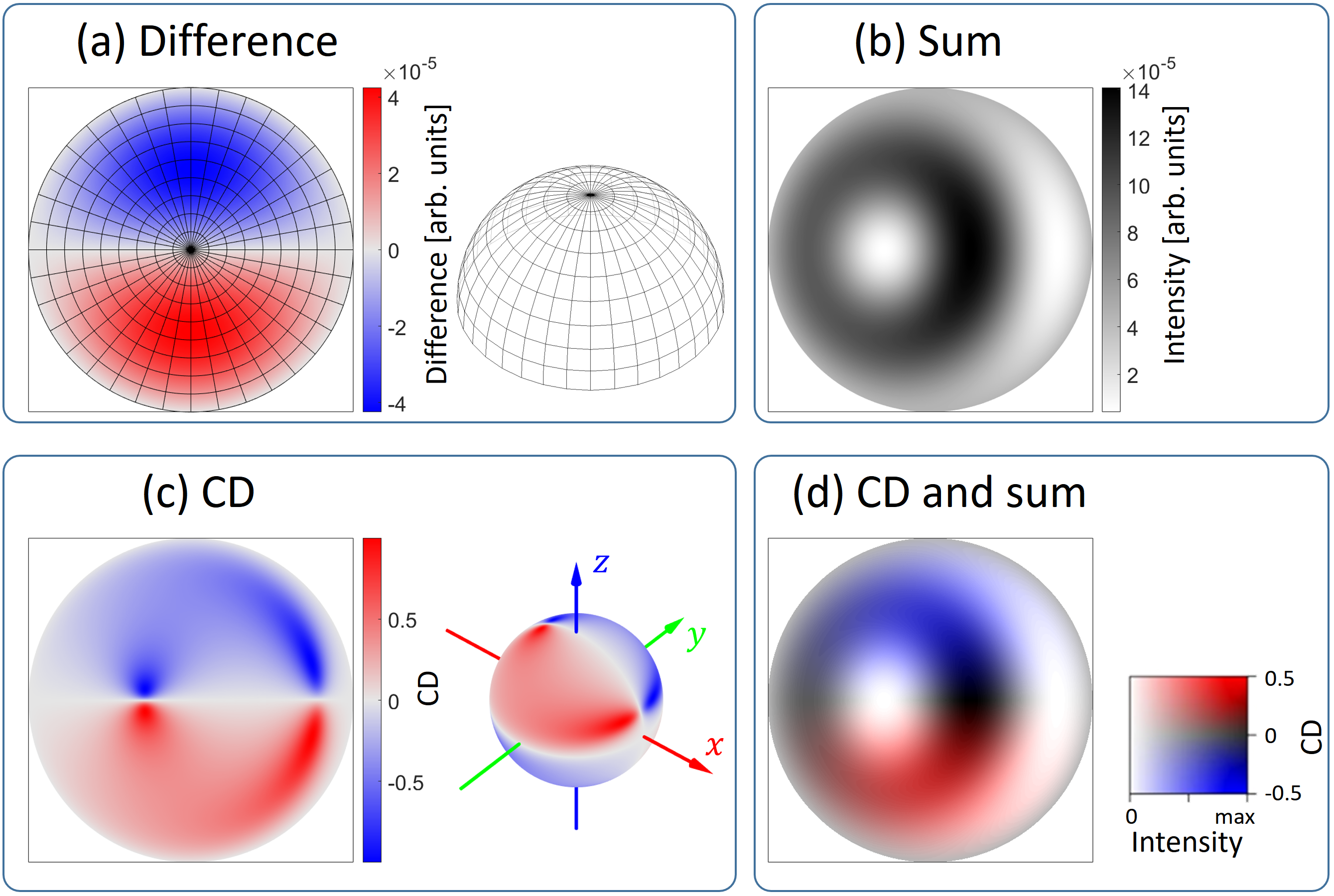}
     \caption{Same as Fig. \ref{fig:CDAD_2pz_80eV} but for $h\nu = 1000$ eV and with the color scale in (d) saturated to $\pm 0.5$.}
     \label{fig:CDAD_2pz_1000eV}
\end{figure}

\section{Photoemission from atomic orbitals with circular light incident at an angle}

It is now possible to calculate the photoemission with $C_\pm$ light from orbitals with quantization axis along the light incidence direction using elementary algebra. The $Y_{l }^{m}( \mathbf{\hat k}_f)$ factor in the partial wave expansion of the scattering state, Eq. \ref{eq:partial_wave}, is the factor outside of the spatial integration on the matrix element. Its physical meaning, for the circular light, is that the angular distribution for any final state orbital will have the shape of the $Y_l^m$ spherical harmonic for that final state orbital, with its quantization axis dictated by the direction of the light incidence. For example, it means that the emission from $Y_1^0 \equiv (1,0)$ orbital with $C_+$ light incident along the $\mathbf z$ axis will be in the form of $Y_2^1\equiv (2,1)$ spherical harmonic, electrons will be primarily emitted along cones at $45^\circ$, around the $\mathbf z$ axis (the $p\rightarrow s$ channel is forbidden in this case). In the following we will use both $Y_l^m$ and $(l,m)$ equivalent notations.

In order to attack the problem with light incidence at a generic angle, Moser \cite{Moser2023}, working within the velocity form of the matrix element, developed a method exploiting {\it vector spherical harmonics}.  This has been done because within the velocity form of the matrix element simple factorization of the matrix element into radial and spherical integrals is not possible. Velocity form of the matrix element might have several advantages, especially the spatial gradient is disconnected from the atomic center definition \cite{Starace1971}. On the other hand, using vector spherical harmonics adds additional complexity, and at least for CDAD from single orbitals, many properties can be obtained using the simpler length form which has been widely used \cite{Bethe1933,Cooper1968,Dubs1985PRL,Goldberg1981,Stohr-Siegmann}. The advantage of this method is in its clarity for some basic examples. A calculation can be performed using basic algebra of angular integrals and tabulated radial integrals and phase shifts \cite{Goldberg1981}.

CDAD from oriented orbitals depends on the light incidence angle. One can account for this by decomposing of the light polarization vector $\boldsymbol \varepsilon$ using $Y_1^{-1}$, $Y_1^0$, and $Y_1^1$ spherical harmonics, the method used e.g. in Refs. \cite{Moser2023,Kern2023}. Here for didactic reasons we will instead decompose $p$ orbitals along the quantization axis defined by the light incidence. Both methods lead to exactly the same results, and in this section we will use the orbital decomposition because it might be more intuitive simple cases. However, in the case of $l \leq 2$ orbitals, the method with decomposing the light polarization is more convenient because it then contains less coefficients. We will use the light polarization decomposition in the section on IMFP-derived dichroism. Furthermore, sometimes the formalism of Wigner $3-j$ symbols in used in this context \cite{Cherepkov1982}, but it will not be used here.

The decomposition of spherical harmonics in the new coordinate system can be made using Wigner $D$-matrices, according to the equation

\begin{equation}
\begin{aligned}
Y_{l,rot}^m = \sum_{m'}d_{mm'}^l ~ Y_l^m
\end{aligned}
\label{eq:Wigner}
\end{equation}

where $m'$ refers to a new quantization axis, and $m$ to the original quantization axis. The formulas for the $D$ symbols for $p$ orbitals are

$d_{1,1}^1 = \frac{1}{2}(1+\cos{\theta}$)

$d_{1,0}^1 = -\frac{1}{\sqrt{2}}\sin{\theta}$

$d_{1,-1}^1 = \frac{1}{2}(1-\cos{\theta}$)

$d_{0,0}^1 = \cos{\theta}$

with $d^l_{m',m} = (-1)^{m-m'}d^l_{m,m'}=d^l_{-m,-m'}$.

However, for $p$ orbitals as initial states, and circular light incidence at $45^\circ$ with respect to the orbital quantization axis, the transformation can be done by elementary algebra, and we provide it for didactic reasons. In the following we set the orbital quantization axis to $\hat z$ and the light incidence direction to $\hat z'$, within the $xz$ plane, see Fig. \ref{fig:pz_rotation}. The angle between $\hat z$ and $\hat z'$ is $45^\circ$.

The $Y_1^0 \equiv (1,0)$ orbital is of particular interest, because it represents the orbital character of graphene Dirac bands. Angular parts of complex $p$ orbitals are conventionally written as

\begin{equation}
\begin{aligned}
(1,-1) = \frac{1}{2} \sqrt{\frac{3}{2\pi}} \cdot e^{-i\varphi}\cdot \sin{\theta}  ~ = ~ & \sqrt{\frac{1}{2}} (p_x - ip_y) \\
(1,0) = -\frac{1}{2} \sqrt{\frac{3}{\pi}} \cdot \cos{\theta} ~ = ~ & p_z \\
(1,1) = -\frac{1}{2} \sqrt{\frac{3}{2\pi}} \cdot e^{i\varphi}\cdot \sin{\theta} ~ = ~ & -\sqrt{\frac{1}{2}} (p_x + ip_y)
\end{aligned}
\end{equation}

where $(l,m)\equiv Y_l^m$. Angular parts of {\it real orbitals} can be written as linear combinations of complex orbitals as
\begin{equation}
\begin{aligned}
p_x = ~ & \sqrt{\frac{1}{2}} \big(~(1,-1) - (1,1) ~\big) \\
p_y = ~ & i \sqrt{\frac{1}{2}} \big(~ (1,-1) + (1,1) ~\big) \\
p_z = ~ & (1,0)
\end{aligned}
\end{equation}

We will rewrite the $p_z = (1,0)$ orbital in the coordinate axis along $\hat z'$. To do that, we first note, that in this coordinate system $p_z = \sqrt{\frac{1}{2}} ( p_{x'} + p_{z'})$, and we also note that in this case $p_{y'} = p_y$. Now we simply need to rewrite this in complex orbitals along $\hat z'$. Using the expressions above we get

\begin{equation}
\begin{aligned}
p_z = \sqrt{\frac{1}{2}} ( p_{x'} + p_{z'}) =  \\ \sqrt{\frac{1}{2}} \Big( \sqrt{\frac{1}{2}} \big(~(1,-1)_{\hat z'} - (1,1)_{\hat z'} ~\big) + (1,0)_{\hat z'} \Big)
\end{aligned}
\end{equation}

and finally

\begin{equation}
\begin{aligned}
p_z  = \frac{1}{2}(1,-1)_{\hat z'} - \frac{1}{2}(1,1)_{\hat z'} + \sqrt{\frac{1}{2}} (1,0)_{\hat z'}
\end{aligned}
\label{eq:pz_decomposition}
\end{equation}

where subscript $\hat z'$ indicates the quantization axis. One can double check with Wigner $D$-symbols for $\theta=-45^\circ$ that $d_{0,-1}^1 = \frac{1}{2}$, $d_{0,1}^1 = -\frac{1}{2}$, and $d_{0,0}^1 = \sqrt{\frac{1}{2}}$. This decomposition is illustrated in Fig. \ref{fig:pz_rotation}, where, with some effort, by visual inspection one can also confirm how the phases of the angular parts of complex orbitals act to obtain the desired decomposition. This form allows applying familiar selection rules without any further considerations. The available channels, with their angular $C_{l\rightarrow l\pm 1}^{m\rightarrow m\pm 1} =  \langle Y_{l\pm 1,m\pm 1} | C_\pm | Y_{lm} \rangle$ and radial $R_{l \rightarrow l \pm 1}$ integrals, angular dependencies $Y_{l\pm 1}^{m\pm 1}( \mathbf{\hat k}_f)$, phases and phase shifts from the partial wave expansion are listed in Table \ref{table} for $C_\pm$ light.

\begin{table}
\begin{tabular}{| l | c | }
\hline \hline
$C_+$ light & \\
\hline
$(1,-1)_{\hat z'} \rightarrow$ &  \\
$l+1$ channel & $i^{1+1}~e^{i\sigma_{1+1}}  ~  C_{l=1\rightarrow 2}^{m=-1\rightarrow 0} ~\cdot R_{1 \rightarrow 2} \cdot Y_2^0( \mathbf{\hat k}_f)$ \\
$l-1$ channel & $ i^{1-1}~e^{i\sigma_{1-1}}    ~ C_{l=1\rightarrow 0}^{m=-1\rightarrow 0} ~\cdot R_{1 \rightarrow 0} \cdot Y_0^0( \mathbf{\hat k}_f)$ \\
\hline
$(1,1)_{\hat z'} \rightarrow$ & \\
$l+1$ channel & $  i^{1+1}~e^{i\sigma_{1+1}}~ C_{l=1\rightarrow 2}^{m=1\rightarrow 2} \cdot R_{1 \rightarrow 2} \cdot Y_2^2( \mathbf{\hat k}_f)$ \\
\hline
$(1,0)_{\hat z'} \rightarrow $ & \\
$l+1$ channel & $ i^{1+1}~e^{i\sigma_{1+1}} ~ C_{l=1\rightarrow 2}^{m=0\rightarrow 1} \cdot R_{1 \rightarrow 2} \cdot Y_2^1( \mathbf{\hat k}_f) $ \\
\hline \hline
$C_-$ light & \\
\hline
$(1,-1)_{\hat z'} \rightarrow$ &  \\
$l+1$ channel & $i^{1+1}~e^{i\sigma_{1+1}}   ~  C_{l=1\rightarrow 2}^{m=-1\rightarrow -2} ~\cdot R_{1 \rightarrow 2} \cdot Y_2^{-2}( \mathbf{\hat k}_f)$ \\
\hline
$(1,1)_{\hat z'} \rightarrow$ & \\
$l+1$ channel & $ - i^{1+1}~e^{i\sigma_{1+1}}  ~ C_{l=1\rightarrow 2}^{m=1\rightarrow 0} \cdot R_{1 \rightarrow 2} \cdot Y_2^0( \mathbf{\hat k}_f)$ \\
$l-1$ channel & $ - i^{1-1}~e^{i\sigma_{1-1}}    ~ C_{l=1\rightarrow 0}^{m=1\rightarrow 0} ~\cdot R_{1 \rightarrow 0} \cdot Y_0^0( \mathbf{\hat k}_f)$ \\
\hline
$(1,0)_{\hat z'} \rightarrow $ & \\
$l+1$ channel & $ i^{1+1}~e^{i\sigma_{1+1}} ~ C_{l=1\rightarrow 2}^{m=0\rightarrow -1} \cdot R_{1 \rightarrow 2} \cdot Y_2^{-1}( \mathbf{\hat k}_f) $ \\
\hline \hline
\end{tabular}
\caption{\label{table} Matrix elements for various $l\pm 1$ channels from the various $l=1$ orbitals. For the $p_z$ orbital with the $45^\circ$ light angle incidence one needs to include the coefficients from Eq. \ref{eq:pz_decomposition} into their respective initial state channels.}
\end{table}

\begin{figure}
 \centering
     \includegraphics[width=8cm]{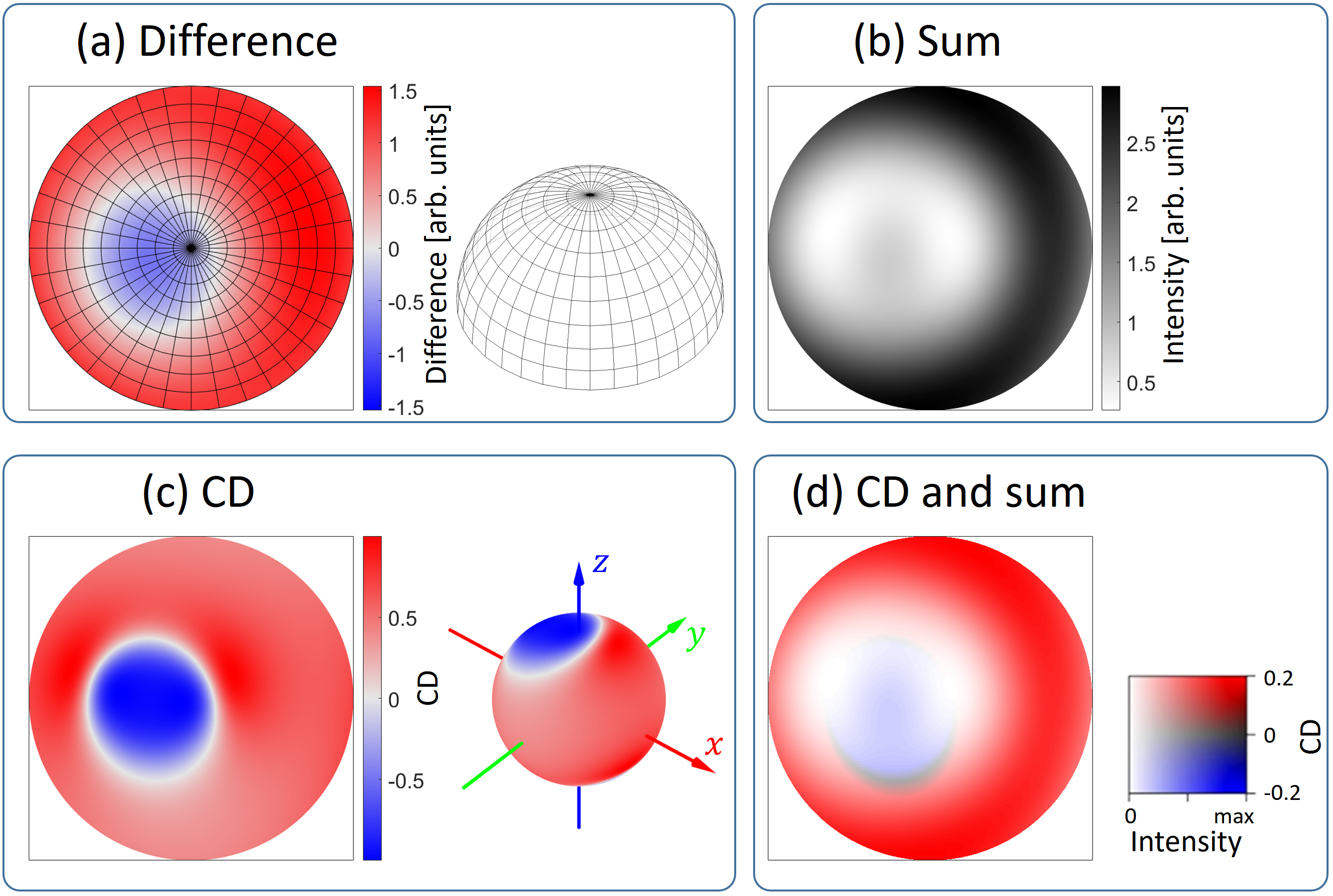}
     \caption{Same as Fig. \ref{fig:CDAD_2pz_80eV} but for $Y_1^1 \equiv (1,1)$ initial state. Photon energy $h\nu = 80$ eV.}
     \label{fig:CDAD_Y1p1_80eV}
\end{figure}

\begin{figure}
 \centering
     \includegraphics[width=8cm]{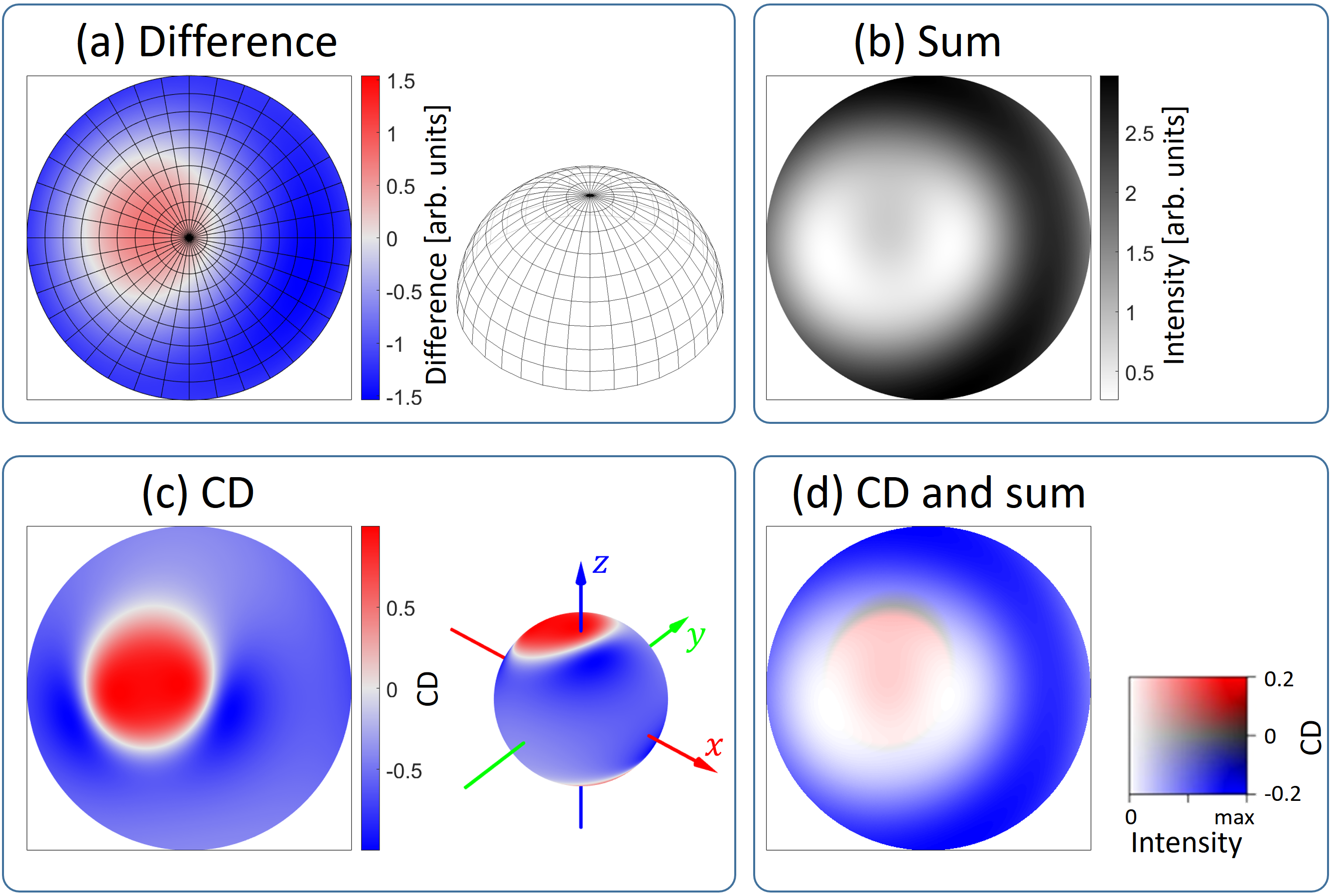}
     \caption{Same as Fig. \ref{fig:CDAD_2pz_80eV} but for $Y_1^{-1} \equiv (1,-1)$ initial state. Photon energy $h\nu = 80$ eV.}
     \label{fig:CDAD_Y1m1_80eV}
\end{figure}

\begin{figure}
 \centering
     \includegraphics[width=8cm]{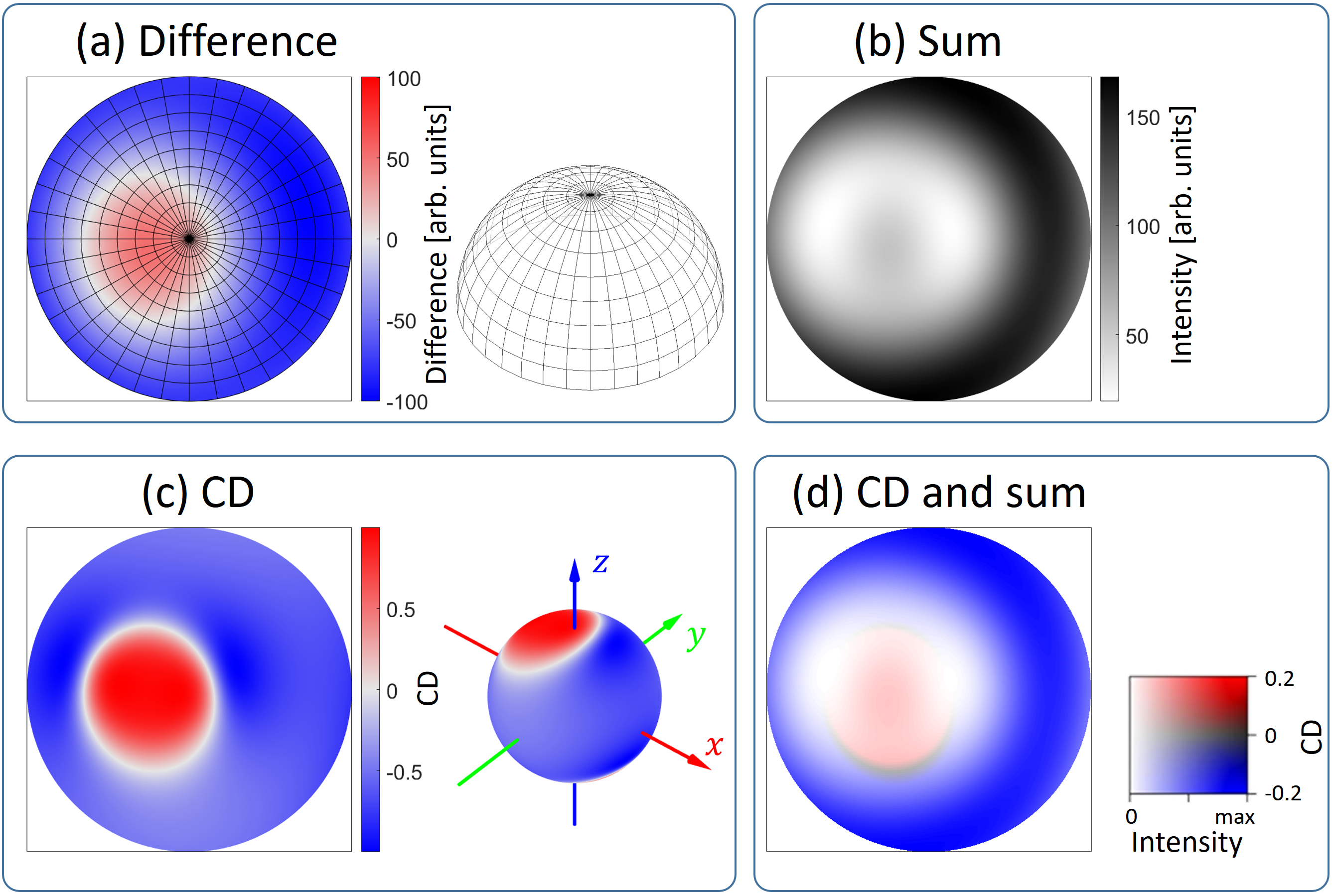}
     \caption{Same as Fig. \ref{fig:CDAD_2pz_80eV} but for $Y_1^{-1} \equiv (1,-1)$ initial state and $h\nu = 21.2$ eV.}
     \label{fig:CDAD_Y1m1_21eV}
\end{figure}

\begin{figure}
 \centering
     \includegraphics[width=8cm]{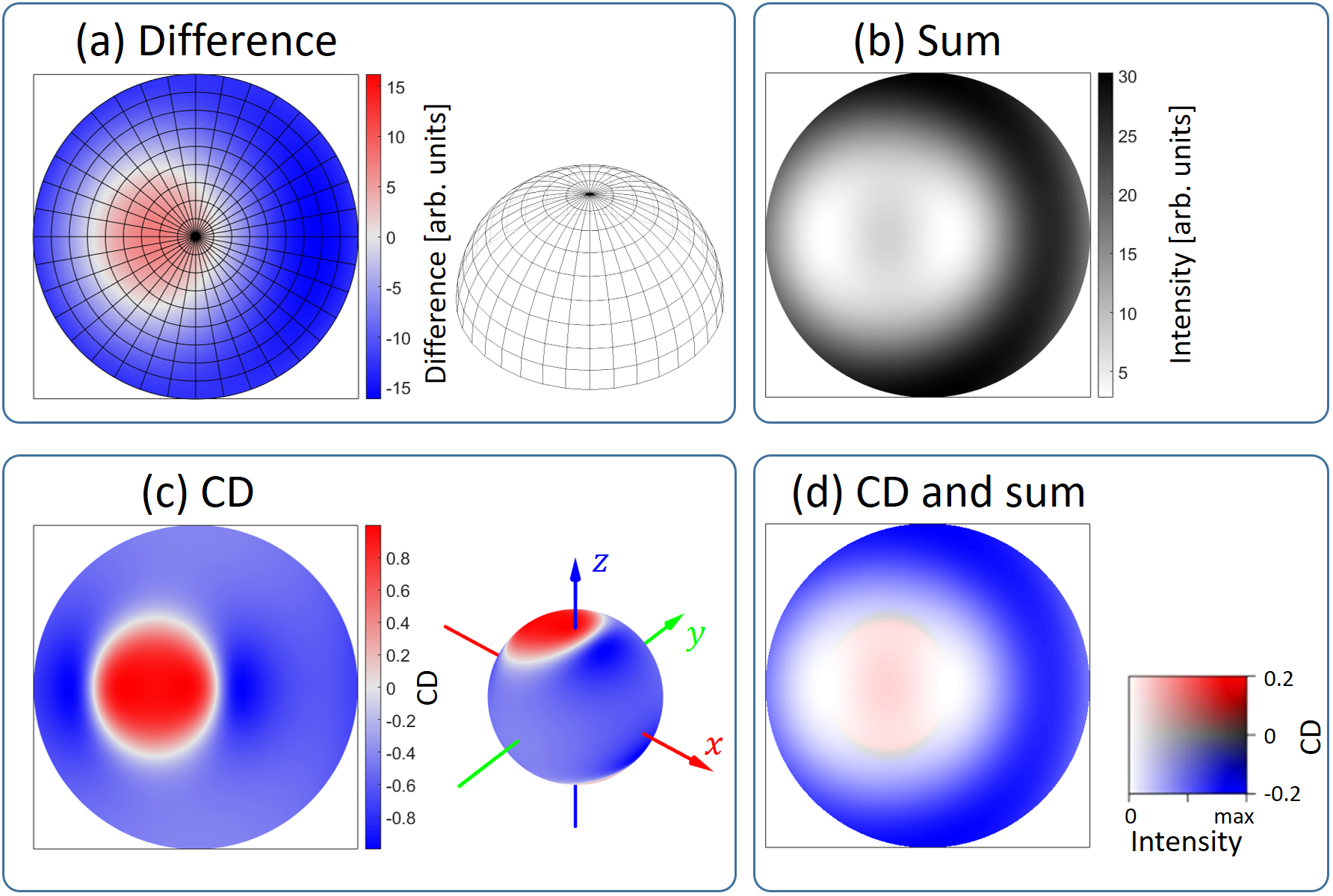}
     \caption{Same as Fig. \ref{fig:CDAD_2pz_80eV} but for $Y_1^{-1} \equiv (1,-1)$ initial state and $h\nu = 40.8$ eV.}
     \label{fig:CDAD_Y1m1_40eV}
\end{figure}

\begin{figure}
 \centering
     \includegraphics[width=8cm]{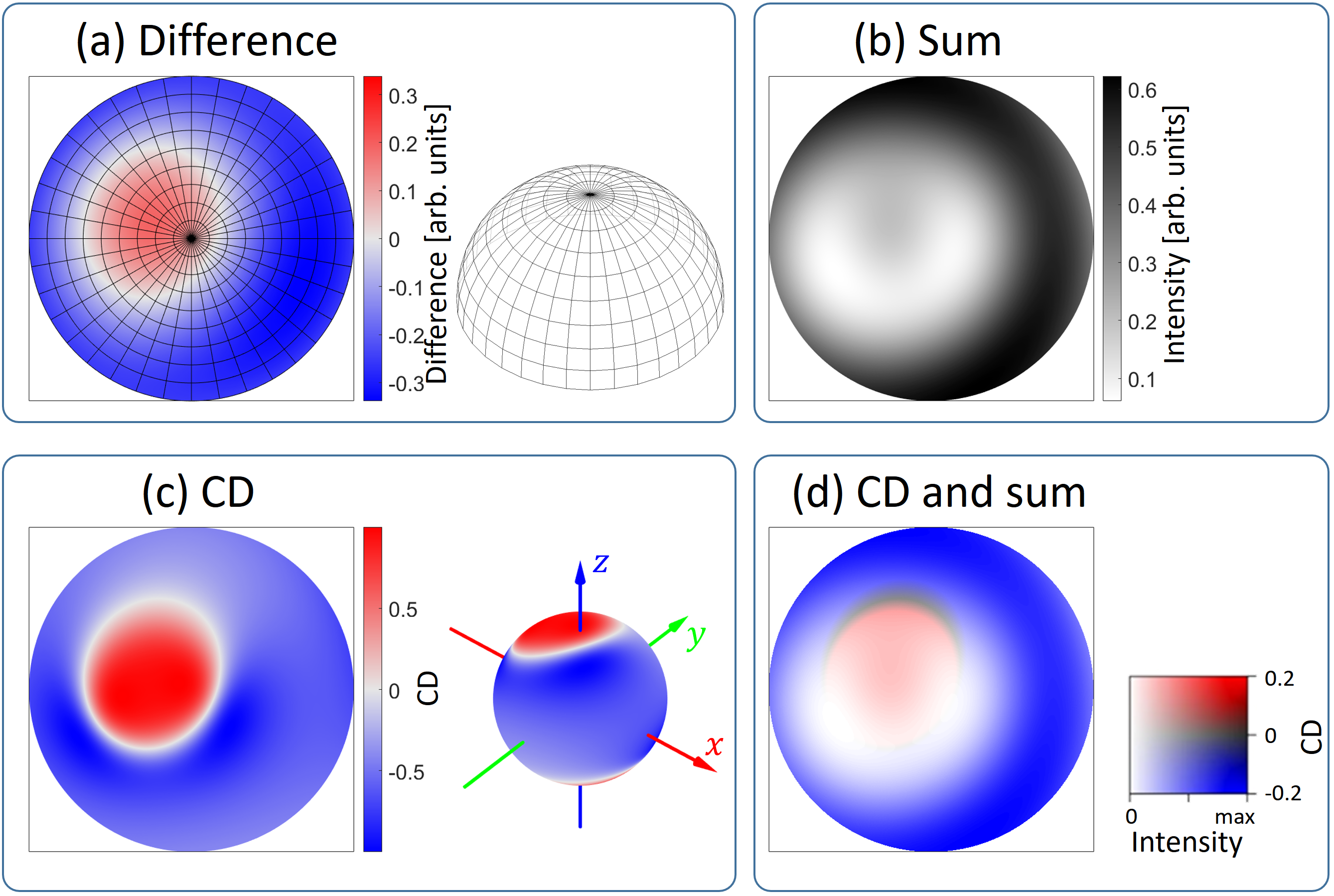}
     \caption{Same as Fig. \ref{fig:CDAD_2pz_80eV} but for $Y_1^{-1} \equiv (1,-1)$ initial state and $h\nu = 120$ eV.}
     \label{fig:CDAD_Y1m1_120eV}
\end{figure}

\begin{figure}
 \centering
     \includegraphics[width=8cm]{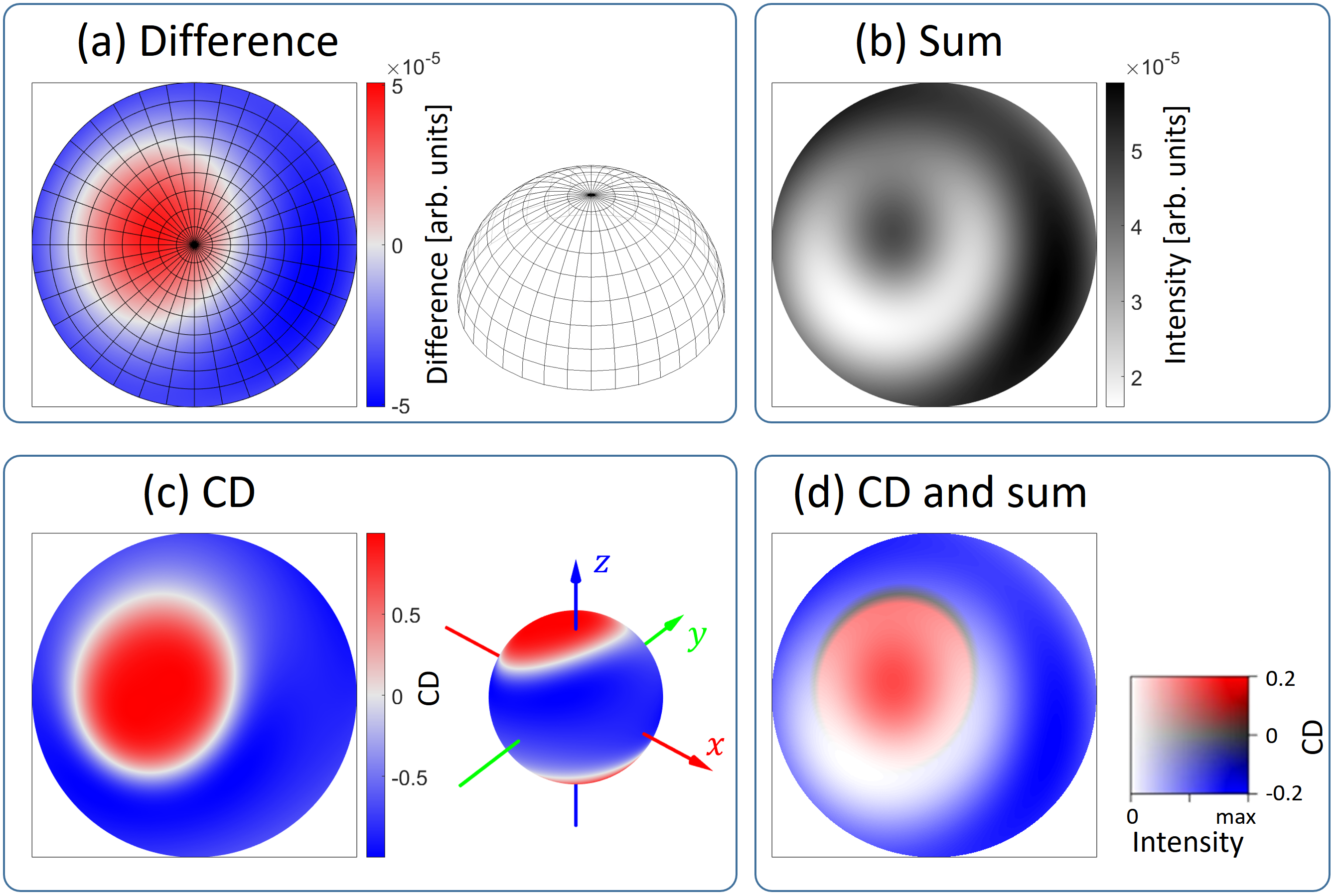}
     \caption{Same as Fig. \ref{fig:CDAD_2pz_80eV} but for $Y_1^{-1} \equiv (1,-1)$ initial state and $h\nu = 1000$ eV.}
     \label{fig:CDAD_Y1m1_1000eV}
\end{figure}

Through the $Y_{l\pm 1}^{m\pm 1}( \mathbf{\hat k}_f)$ terms one gets the angular dependence of the photocurrent. In order to calculate the photocurrent, one needs to sum all the components, and square the absolute value of the sum. This way one can check that the CDAD from $Y_1^0 \equiv (1,0)$ orbital will be non zero only if the Coulomb phase shifts differ, and the phase shifts can differ only between the $l+1$ and $l-1$ channels. Varying the ratios of the radial integrals $R_{l\pm 1}$, while keeping the phase shifts the same for $l\pm 1$ channels, will not lead to a non-zero CDAD signal from the $p_z$ orbital.


Figure \ref{fig:CDAD_2pz_80eV} shows the calculated CD-ARPES signals from the $Y_1^0 \equiv (1,0)$ orbital with light incident at $45^\circ$, using C $|2p\rangle$ parameters from Goldberg et al. \cite{Goldberg1981}, which are also plotted in Fig. \ref{fig:Goldberg}. The 3D visualization in Fig. \ref{fig:CDAD_2pz_80eV}(c) shows that CDAD vanishes within $xy$ and $xz$ planes. Detailed conditions for CDAD with respect to the direction of the incoming light are given in \cite{Dubs1985}, essentially the CDAD will vanish if $C_\pm$ light is incident along the quantization axis of the $p_z$ orbital. 

Figure \ref{fig:CDAD_2pz_21eV} shows the same as Fig. \ref{fig:CDAD_2pz_80eV} but for $h\nu = 21.2$ eV. One can see that the sign of CDAD signal reverses. This is the result of the change in the phase shift difference, which crosses the $-180^\circ$ at around $h\nu = 40$ eV, as tabulated in Ref. \cite{Goldberg1981} and shown in Fig. \ref{fig:Goldberg}.

Figure \ref{fig:CDAD_2pz_40eV} shows the same as Fig. \ref{fig:CDAD_2pz_80eV} but for $h\nu = 40.8$ eV. One can see that the CDAD signal is very weak. This is the result of the change in the phase shift difference, being close to $-180^\circ = -\pi$ for $h\nu = 40.8$ eV, as tabulated in Ref. \cite{Goldberg1981} and shown in Fig. \ref{fig:Goldberg}. The phase shift of $-180^\circ$ means a sign change. Therefore the CD-ARPES signal due to the Coulomb phase shift difference mechanism can be expected to vanish near the Fermi level of graphene and graphite for $h\nu \approx 40$ eV.

Figure \ref{fig:CDAD_2pz_1000eV} shows the same as Fig. \ref{fig:CDAD_2pz_80eV} but for $h\nu = 1000$ eV. Clearly, the total cross section in Fig. \ref{fig:CDAD_2pz_1000eV}(b) becomes much smaller, however, the CDAD signal becomes slightly stronger than for the VUV energy range. This shows that CDAD from C $|2p_z\rangle$ remains strong over the entire VUV and soft x-ray photon energy range (excluding the $h\nu \approx 40$ eV region).

All the CDAD maps for the $p_z$ initial state obey the axial vector rules for the $\mathcal M_y$ (i.e. $xz$) mirror plane. This is because the initial $p_z$ orbital wave function obeys the $\mathcal M_y$, and the the object $I_{C_+} - I_{C_-}$ does not break the $\mathcal M_y$ symmetry, because $\mathcal M_y$ operation turns the $C_+$ light into the $C_-$ light while keeping the propagation direction. Light propagation is parallel to the $\mathcal M_y$ mirror plane, the CDAD signal must reverse on the other side of the mirror plane, according to the axial vector mirror reflection rules.

The formalism sketched above allows to derive CD-ARPES pattern from any orbital using the length form of the matrix element. Importantly, the standard derivation of the CDAD signal from the $m\neq 0$ orbitals has been routinely based on the ratio of radial integrals for $l\pm 1$ channels, and does not vanish even without taking into account the Coulomb phase shifts. However, the phase shifts will additionally contribute also to the CDAD from $m\neq 0$ orbitals. Therefore, we will also derive the CDAD signal for the $Y_1^{\pm 1} \equiv (1, \pm 1)$ orbitals. One can again decompose the orbital using geometrical arguments, but this time we will just use  Wigner $D$-symbols for brevity. We can rewrite $(1,1)$ along the $\hat z'$ quantization axis as

\begin{equation}
\begin{aligned}
(1,1)_{\hat z}  =
d_{1,-1}^1 (1,-1)_{\hat z'} + d_{1,1}^1 (1,1)_{\hat z'} + d_{1,0}^1 (1,0)_{\hat z'}
\end{aligned}
\label{eq:1p1_decomposition}
\end{equation}

where for $-45^\circ$ incidence we have $d_{1,-1}^1 = \frac{1}{2} \bigg(1- \frac{1}{\sqrt 2}\bigg)$, $d_{1,1}^1 = \frac{1}{2}\bigg(1+ \frac{1}{\sqrt 2}\bigg)$, and $d_{1,0}^1 = 0.5$.


Now we can use the same expressions as in Table \ref{table} but with the coefficients above, with the results shown in Figs. \ref{fig:CDAD_Y1p1_80eV},
\ref{fig:CDAD_Y1m1_80eV}, \ref{fig:CDAD_Y1m1_21eV}, \ref{fig:CDAD_Y1m1_40eV}, \ref{fig:CDAD_Y1m1_120eV}, and \ref{fig:CDAD_Y1m1_1000eV}. We can see, that unlike the results for the $p_z$ orbital, these figures do not obey the $\mathcal M_y$ axial vector symmetry. This is because the initial state orbitals $Y_1^{\pm 1} \equiv (1,\pm1)$ are not mirror symmetric with respect to $\mathcal M_y$. However, comparing Figs. \ref{fig:CDAD_Y1p1_80eV} and \ref{fig:CDAD_Y1m1_80eV} one can see that the two CDAD maps are connected to each other by the $\mathcal M_y$ operation (taking into account axial vector rules which reverse colors). This is because, up to the arbitrary phase factor, $\mathcal M_y Y_1^1 = Y_1^{-1}$. We can also see that the result for $h\nu = 40.8$ eV is more symmetric (albeit not in accord to the $\mathcal M_y$ axial rules, since the $\mathcal M_y$ symmetry is broken), due to the Coulomb phase shift difference being $\approx -180^\circ$.

\section{Dichroism due to the inelastic mean free path}

\begin{figure}
 \centering
     \includegraphics[width=6cm]{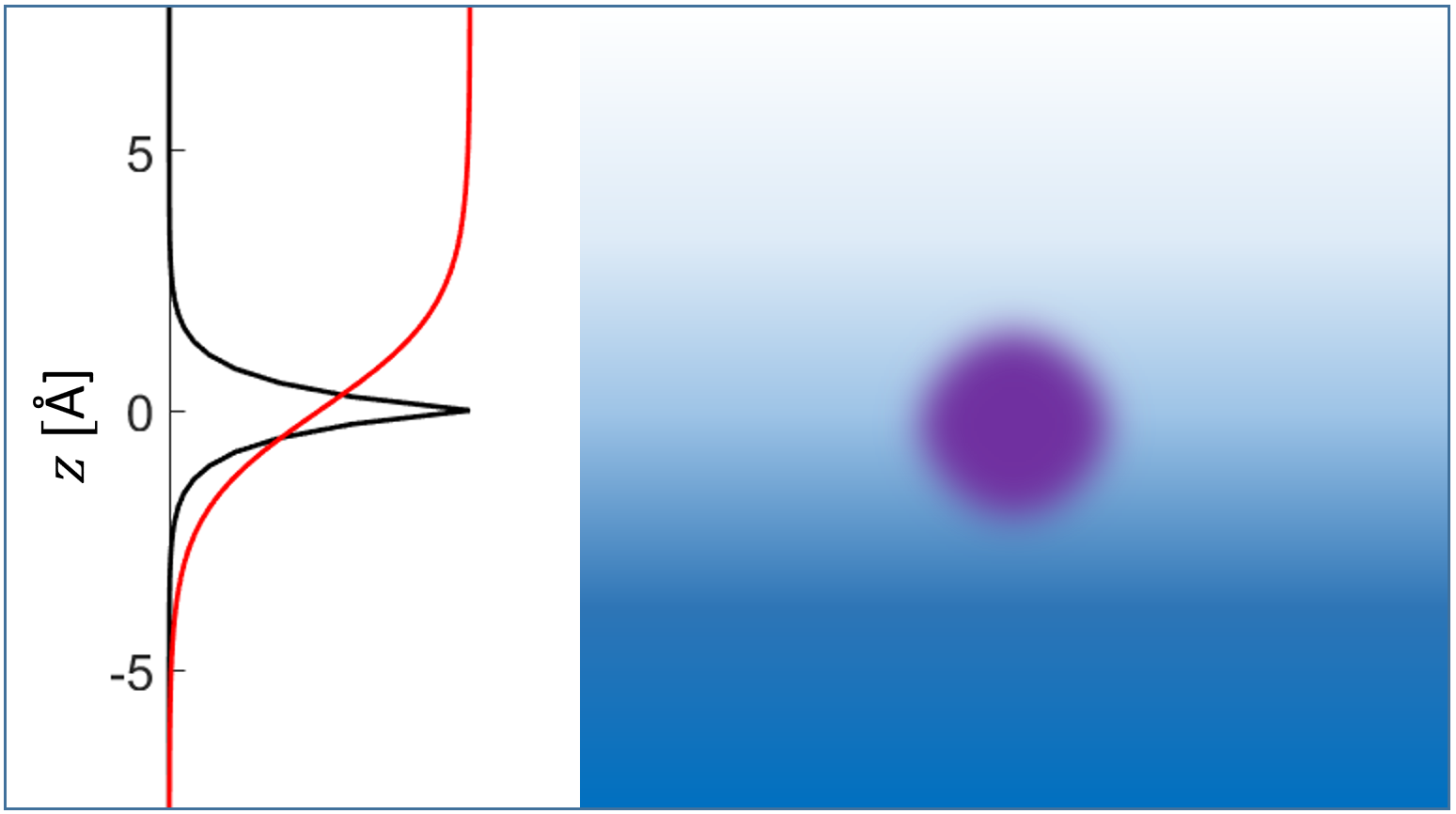}
     \caption{Schematic illustration of the $|1s\rangle$ orbital submerged in the sea of electronic gas near the surface of a material. The left panel shows the radial component of the H $|1s\rangle$ wave function (black curve) together with the Fermi distribution function $P(z) = (e ^{z/\lambda}+1)^{-1}$ for $\lambda = 1$ \AA, which can be interpreted as corresponding to the IMFP of $\approx 4\cdot\lambda = 4$ \AA.}
     \label{fig:IMFP_schematic}
\end{figure}

Emission of electrons from solids is always influenced by the IMFP for the electrons moving inside the solid. Moser \cite{Moser2017} has established that IMFP-derived CD-ARPES signal is present already within the free-electron final state (FEFS) approximation and in the following we will derive this process by considering an orbital submerged in the electron gas near the surface of the material, as illustrated in Fig. \ref{fig:IMFP_schematic}. The influence of mean free path is typically modeled through the exponential function as $P(z) = e^{-z/\lambda}$, where $z$ is the distance to the surface, and $\lambda$ is a decay parameter proportional to the IMFP. This description is artificial because it requires the function $P(z)$ to be truncated at the surface, and on the atomic scale it not possible to define an exact position of the surface. Therefore, in general the function $P(z)$ is not known, and for the electron-electron scattering mechanism one could imagine to derive it from the charge density of the material. The matrix element for the emission of electrons from the solid can be then written with the inclusion of the IMFP by including it as a factor inside the space integral

\begin{equation}
\begin{aligned}
M_{fi} = \int P(z)~ \psi_f \cdot \boldsymbol \varepsilon \cdot \mathbf r \cdot \psi_i ~ d\mathbf r
\end{aligned}
\end{equation}

where we have used the length form of the matrix element. When using FEFS, $\psi_f = e^{i\mathbf k_f \cdot \mathbf r}$, it is convenient to calculate this integral numerically on the 3D mesh, since the initial hydrogenic orbital wave functions $\psi_i = \phi_{nlm}(\mathbf r)$ are easily available. In the following we will ignore the amplitude $|\varepsilon|$. Circularly polarized light $C_\pm$ incident along the $z$ axis is expressed as $\boldsymbol \varepsilon \propto \varepsilon_x \pm i \varepsilon_y$ where we ignore the $\sqrt{2}$ normalization factor for convenience. Here, e.g. by $\varepsilon_x$ we mean the vector $[\varepsilon_x, 0, 0]$. For a generic light incidence angle we either need to include the IMFP profile at an angle with respect to the light incidence, or rewrite the expression for light polarization using the Wigner $D$-matrix coefficients, and we will do the latter here for didactic reasons. Both methods must lead to the same CDAD profile, which can serve as a test for possible issues with the numerical integration. Rewriting the light polarization vector in the new coordinate system follows the same principle as the rewriting the $p$ orbitals along the new quantization axis, and it has a meaning of expressing $\boldsymbol \varepsilon$ in $l=1$ spherical harmonics. Just like for the decomposition in Fig. \ref{fig:pz_rotation}, for the $45^\circ$ angle of light incidence we have $\boldsymbol \varepsilon \propto (\varepsilon_x + \varepsilon_z)/\sqrt{2} \pm i \varepsilon_y$. In this case the matrix element has the form

\begin{equation}
\begin{aligned}
M_{fi,C_\pm} = \int P(z)~ e^{i\mathbf k_f \cdot \mathbf r} \cdot \big( (\varepsilon_x + \varepsilon_z)/\sqrt{2} \pm i \varepsilon_y \big) \cdot \mathbf r \cdot \psi_i ~ d\mathbf r
\end{aligned}
\label{eq:IMFP}
\end{equation}


In general, the form of the function $P(z)$ is not known, however, the boundary conditions should be, that it rapidly becomes zero few nanometers below the surface, and unity few nanometers above the surface. Therefore, the Fermi function $P(z) = (e ^{z/\lambda}+1)^{-1}$ is a good candidate, where IMFP can be approximately taken as $4 \lambda$. We illustrate such Fermi function IMFP profile over the H $|1s\rangle$ radial wave function in Fig. \ref{fig:IMFP_schematic}.

\begin{figure}
 \centering
     \includegraphics[width=8cm]{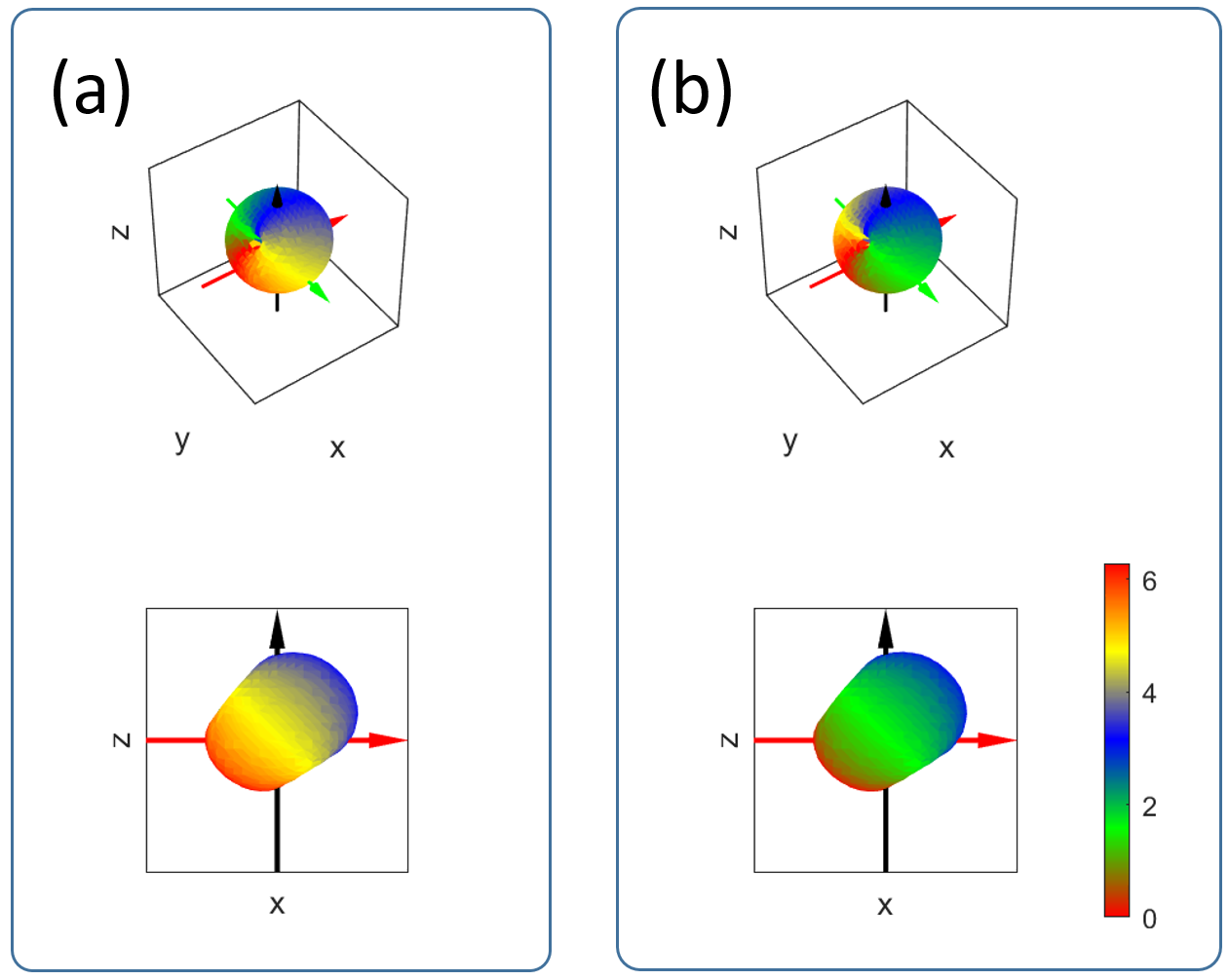}
     \caption{The $P(z) ~ \boldsymbol \varepsilon \cdot \mathbf r |\psi_i \rangle$ object for $\psi_i = |1s\rangle$ of hydrogen. This object enters the length form of the matrix element. (a) 3D and side view isosurface for $C_+$ light. (b) The same for $C_-$ light. Light is incident with the $yz$ plane at $45^\circ$. Colormap shows the complex phase between $0$ and $2\pi$, and it is shown in the bottom right.}
     \label{fig:objectIMFP}
\end{figure}

\begin{figure}
 \centering
     \includegraphics[width=8cm]{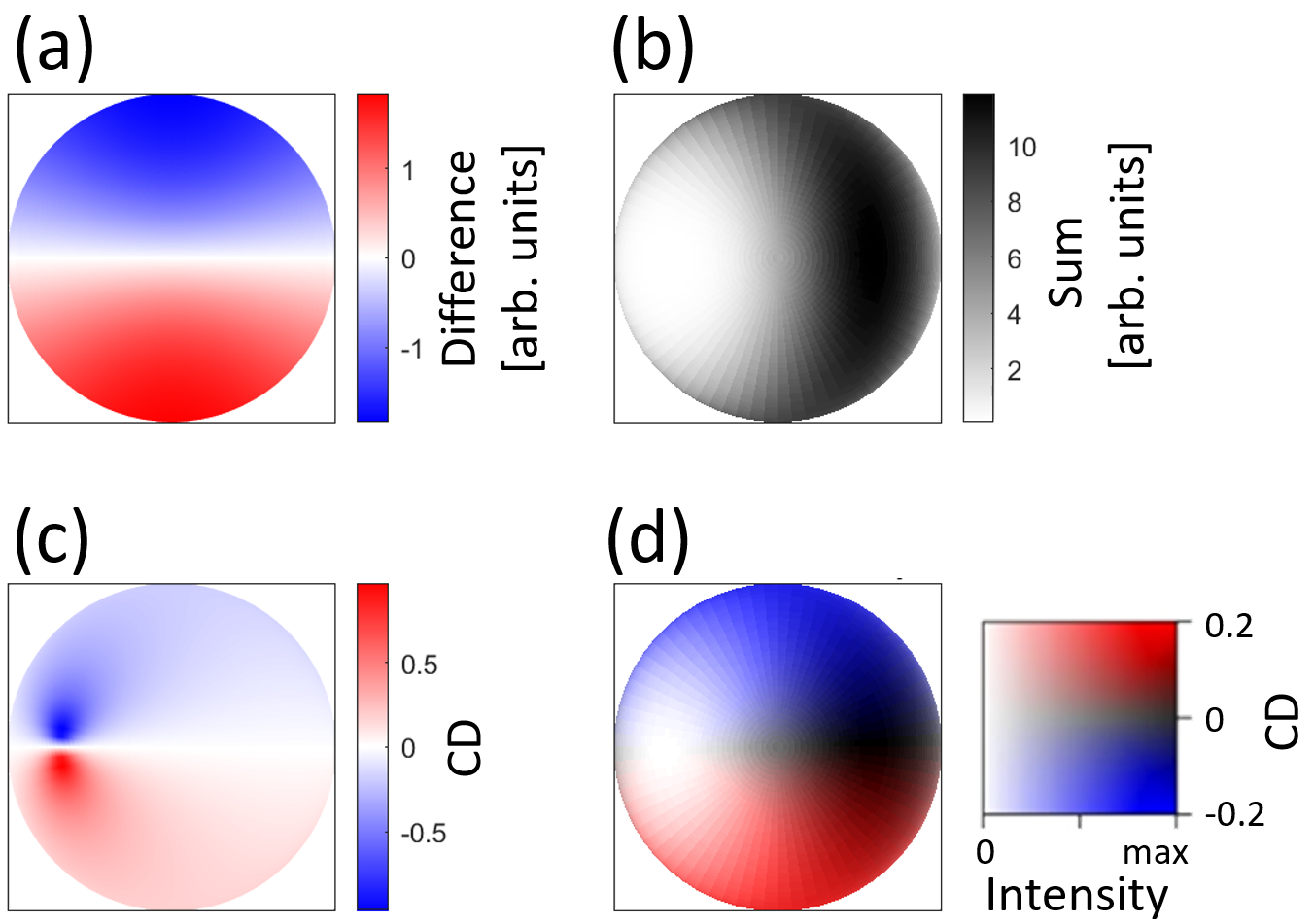}
     \caption{Numerical calculation of the CD signal from and the hydrogen $|1s\rangle$ orbital assuming the Fermi function IMFP profile, the length form of the matrix element, and the FEFS at $h\nu = 50$ eV. (a) Difference between the intensities with $C_+$ and $C_-$ light. (b) Sum of these intensities (c) Dichroism, difference divided by the sum. (d) Same as (c) but using the 2D colormap shown in bottom right with the color scale saturated at $20\%$. }
     \label{fig:IMFP}
\end{figure}

Let us focus on the initial wavefunction being H $|1s\rangle$ orbital, $\psi_i = \phi_{100}$. First we consider the object

\begin{equation}
\begin{aligned}
\boldsymbol \varepsilon \cdot \mathbf r |s \rangle = \big( (\varepsilon_x + \varepsilon_z)/\sqrt{2} \pm i \varepsilon_y \big) \cdot \mathbf r \cdot \phi_{100}
\end{aligned}
\label{eq:object}
\end{equation}

This object  must have an angular dependence of the $(1,1)_{z'}$, quantized along the light incidence $z'$. This is a result of the partial wave expansion of the plane wave Eq. \ref{eq:partial_wave_FEFS}, where, from the selection rules, we know that only the final state $Y_1^1$ is available with $C_+$ light from the $Y_0^0$ initial orbital. Getting more into details, it is a result of the $Y_l^{m*}(\mathbf {\hat r}) Y_l^m(\mathbf{\hat k}_f)$ factor in the partial wave expansion, which ensures the same angular dependence of the matrix element as the angular part of the allowed final state orbital.

In order to illustrate the origin of the IMFP-dervied CDAD signal, we can evaluate the product of the function $P(z)$ with the object from the Eq. \ref{eq:object}

\begin{equation}
\begin{aligned}
P(z) ~ \boldsymbol \varepsilon \cdot \mathbf r |s \rangle = P(z) ~ \big( (\varepsilon_x + \varepsilon_z)/\sqrt{2} \pm i \varepsilon_y \big) \cdot \mathbf r \cdot \phi_{100}
\end{aligned}
\label{eq:objectIMFP}
\end{equation}

This object is shown in Fig. \ref{fig:objectIMFP} and it has a shape of a deformed $(1,1)_{z'}$. Effectively, it means that many $Y_l^m$ components will be necessary to decompose it, and therefore it will allow many final state channels in the matrix element. This is the reason why adding IMFP profile can, at least in principle, result in a non-vanishing CDAD even in the FEFS approximation. It is not immediately clear if an analytical calculation can be performed in this case, therefore to check whether the effect indeed takes place, it is convenient to perform numerical calculations.

The result of the numerical integration of Eq. \ref{eq:IMFP} for the initial state $|1s\rangle$ hydrogen orbital at $h\nu = 50$ eV is shown in Fig. \ref{fig:IMFP}. We performed brute-force integration on the $N\times N \times N$ mesh with $N=100$ result shown in Fig. \ref{fig:IMFP}, with the mesh step of 0.5 Bohr. Computations with larger $N$ lead to virtually identical results. For IMFP of $\approx 4$~\AA~ the CD signal of $\approx 20 \%$ is present, and it has a similar character as the CD signal due to Coulomb phase shifts. Since IMFP gets larger at higher kinetic energies of emitted electrons, the IMFP-derived CD-ARPES will vanish at high photon energies, where IMPF is large (see also Ref. \cite{Moser2017}). We note that the absolute sign of this dichroic signal relative to the Coulomb-phase-shift derived signal is not unambiguous here and needs to be carefully established.


\section{Discussion and outlook}


This study demonstrates how to calculate CDAD pattern from oriented orbitals using elementary algebra (Coulomb scattering state derived patterns) or numerical integration (IMFP-derived patterns). For the derivations using the Coulomb scattering state, we focus on $p$ orbitals as a simplest non-trivial cases, while for the IMFP-derived CDAD we consider H $|1s\rangle$ orbital. In crystals, multiple scattering will further complicate the appearance of the CD-ARPES patterns, possibly leading to local sign reversals.

In the Kubo formula for quantum transport, transverse conductivity is proportional to the integral of the Berry curvature throughout the Brillouin zone. In general, Berry curvature has large values in the band structure regions where band characters are being reversed and band character is mixed. A classic example is a spinless two-band system with one $m=0$ and one $m=1$ orbital \cite{Nagaosa2010RMP,Xiao2010RMP}. One possible feature of CD-ARPES is that through the relation to the momentum-resolved OAM it may allow experimental access to the Berry curvature or at least to some of its properties \cite{Schueler2020}. Unfortunately, it is clear that strong extrinsic effects in CD-ARPES discussed here are unlikely to be related to Berry curvature physics since many of them are not related to the OAM of the initial orbitals. It therefore important to understand and possibly eliminate, minimize, or perhaps make use of these effects in future CD-ARPES studies. A general path to obtain the connection to the initial OAM band character is through comprehensive analysis of numerous CD-APRES maps taken at different photon energies and different light incidence angles, through comparison to the one-step model calculations \cite{Beaulieu2020}.

In many solid state systems, bands are split according to the orbital characters quantized along the surface normal. Since many effects discussed here vanish for the light incidence along the quantization axis of the orbital, one path to eliminate non-OAM contributions to CD-ARPES would be through experiments performed at normal light incidence. Unfortunately, such experiments are not routinely feasible, due to the current designs of the photoelectron spectrometers. One can imagine bringing the photon beam through the spectrometer lens (such designs have existed in the past \cite{Schneider1990}), however, any scattered light will produce unwanted electron background, and prolonged illumination of spectrometer slits would likely permanently compromise their performance. Nevertheless, perhaps future spectrometer designs can mitigate these issues since no fundamental obstacles exist.

Regarding available designs, when using momentum microscopes sample must be normal to the analyzer lens, which makes normal incidence measurements effectively not possible. For hemispherical analyzers, normal light incidence is in general possible (at many setups the light-lens angle is either $45^\circ$ or $54.7^\circ$) even though it will mean that only electrons emitted at high emission angles will be accessible, which means there is no access to the center of the Brillouin zone. However, in this way, one should be able to access some of the $K$ points of graphene to test of CD-ARPES signal vanishes at normal incidence.

Furthermore, at generic light incidence angle (e.g. $45^\circ$) one shall check if CD-ARPES vanishes near $h\nu \approx 40$ eV, as predicted in the Coulomb scattering state model. Therefore, such regular experiment at a generic light incidence angle would allow to disentangle the IMFP contribution to CDAD signal, and possibly establish how multiple scattering acts in case of these two contributions to CD-ARPES. One difference between the Coulomb-phase-shift and IMFP-derived CDAD for C $|2p_z\rangle$ is that the latter does not vanish for the very large emission angles (i.e. emission parallel to the surface). This can be another test for the contribution of the two discussed processes to the CD-ARPES signal in graphene/graphite.

Importantly, dichroic patterns similar to those from $p$ orbitals can be obtained for the $d_{z^2}\equiv (2,0)$ orbitals with the light incicence at an angle, and for other $l=2$ orbitals. This can be used to explain large CD-ARPES signal in the vicinity of the Brillouin zone center of WSe$_2$ which is derived from W $5d_{z^2}$, and to understand the details of the CD-ARPES texture near the $K/K'$ points which are composed from $Y_2^{\pm2}$ orbitals. Inspection of the radial integrals and phase shifts for W $5d$ \cite{Goldberg1981} reveals vanishing $l+1$ radial matrix element in the region between $h\nu=160$ and 200 eV, and vanishing phase shift difference for $h\nu \approx 100$ eV. Experimental CD-ARPES near $\Gamma$ in WSe$_2$ is strong at a nearby energy $h\nu = 94$ eV \cite{Cho2021}, therefore this CD signal can either orginate primiarily from the IMFP, or from the admixture of the Se $|4p_z\rangle$ orbitals near $\Gamma$. Another possibility is that the phase shift and radial integral calculations need to be revised for the case of delocalized valence orbitals in solids.

Finally, we note that the relation between the total OAM along a certain quantization axis and the CDAD measurement is in general non-trivial for $l>1$ orbitals. With $l=1$ orbitals ($p$ orbitals) the only choice for $m$ is $-1$, $0$, or $1$. These $m$ values are the expectation values of the $L_z$ operator. This means that any CDAD from $p$ orbitals quantized along the light incidence direction $z$ is related to $m\pm 1$ and at least for some emission angles may predictably scale with the expectation value of the $L_z$ operator. However, with $l=2$ orbitals, we have more choices with $m$ from $-2$ to $2$. This makes the connection of the CDAD signal to the $L_z$ expectation value more complicated, because the angular distribution signal from $Y_2^{\pm 1}$ is different from that from $Y_2^{\pm 2}$ due to different available final state channels, and the relation between CDAD and the expectation value of $L_z$ becomes non-trivial.

In summary, this manuscripts discusses CDAD processes from oriented orbitals which are likely responsible for strong CD-ARPES signal in graphene/graphite and in WSe$_2$ family of materials, as well as for CD-ARPES in general. We reproduced previous results on CDAD from oriented orbitals using the Coulomb scattering state method recently revived by Moser \cite{Moser2023}, but within the length form of the matrix element. We also demonstrated the microscopic origin of the IMFP-derived CD-ARPES signal. We provide numerous plots of the expected CDAD, and offer s description of how cases for any other orbital can be computed elementarily using tabulated values and algebraic expressions. One possible future direction would be to establish how multiple scattering and interatomic interference can modify the orbital-derived CD-ARPES signal.



\section{Addendum: Checking how CDAD can be zero with FEFS for circular light incident along the quantization axis of the orbitals}

It is intuitive to assume that circular dichroism from $m\neq 0$ orbital will exist and indeed it does exist in real atoms and solids. However, in case of the FEFS it is easy to show, e.g. through the numerical integration or model considerations \cite{Moser2023}, that this is not the case, and neither the matrix element $\langle e^{i\mathbf{k}\cdot \mathbf{r}}| \boldsymbol \varepsilon \cdot \mathbf r | \phi_{nlm}\rangle$, nor the $\langle e^{i\mathbf{k}\cdot \mathbf{r}}| \mathbf A \cdot \mathbf p | \phi_{nlm}\rangle$, will allow for a non-zero CD-ARPES signal.


With FEFS one can use the partial wave expansion

\begin{equation}
\begin{aligned}
& e^{i\mathbf{k}\cdot \mathbf{r}} = \\
& 4\pi \sum_{l=0}^\infty \sum_{m=-l}^l i^l  j_l(k_f r)~Y_l^{m*}(\mathbf {\hat r}) ~ Y_l^m(\mathbf{\hat k}_f)
\label{eq:partial_wave_FEFS}
\end{aligned}
\end{equation}

with spherical Bessel functions $j_l$ that depend on position $r = |\mathbf r|$ and wavevector amplitude $k = |\mathbf k|$, and therefore also depend on kinetic energy. One can see that this equation is a simplified form of Eq. \ref{eq:partial_wave} without the Coloumb phase shifts, there is only a sign change between $l\pm 1$ channels due to the $i^l$ factor.

With initial channel $Y_0^0 \equiv (0,0)$ CDAD is obviously zero. This is because there are two final state channels $(1,-1)$ and $(1,1)$ respectively for $C_-$ and $C_+$ light. Upon squaring the intensities are the same.

With initial channel $Y_1^1 \equiv (1,1)$ we have final states $(2,0)$ and $(0,0)$ for $C_-$ light and final state $(2,2)$ for $C_+$ light. We write explicitly

$(2,2) = \frac{1}{4} \sqrt{\frac{15}{2\pi}} \cdot e^{2i\varphi}\cdot \sin^2{\theta}$

$(2,0) = \frac{1}{4} \sqrt{\frac{5}{\pi}} \cdot (3\cos^2{\theta}-1)$

$(0,0) = \frac{1}{2} \sqrt{\frac{1}{\pi}}$

The question is if we can get the same angular dependence as the $(2,2)$ has, by combining $(2,0)$ and $(0,0)$. The exponential factor $e^{2i\varphi}$ in $(2,2)$ will square out to unity and is not relevant. One can verify that the $\sin^2{\theta}$ dependence is obtained through

$(2,0)+\sqrt{5} \cdot (0,0) \propto \sin^2{\theta} $

Vanishing CDAD can be then obtained through proper intensity ratios of the $C_+$ and $C_-$ channels. One can check that  these numbers are the same as the angular integrals listed in Table \ref{table:Stohr}. In these sense these angular integrals for the length form of the dipole element are the numbers that give zero CDAD in the FEFS model.

Now let us consider the $Y_2^2 \equiv (2,2)$ initial state. The channels are $(3,3)$ for the $C_+$ light and a linear combination of $(3,1)$ and $(1,1)$ for the $C_-$ light. The relevant $Y_l^m$ functions are

$(3,3) = -\frac{1}{8} \sqrt{\frac{35}{\pi}} \cdot e^{3i\varphi}\cdot \sin^3{\theta}$

$(3,1) = -\frac{1}{8} \sqrt{\frac{21}{\pi}} \cdot e^{i\varphi}\cdot \sin{\theta} \cdot (5\cos^2{\theta}-1)$

$(1,1) = -\frac{1}{2} \sqrt{\frac{3}{2\pi}} \cdot e^{i\varphi}\cdot \sin{\theta}$

The first thing that we notice is that the $\varphi$ dependence is the same in the $p$ and $f$ channels for $C_-$ light, as it actually has to be through the selection rules. This allows to factorize the exponential oscillating factor that will square to unity. Then it is easy to calculate that

$(3,1)-\sqrt{14}\cdot (1,1) \propto \sin^3{\theta}$

Therefore, we conclude, that at least in the length form of the matrix element, the CDAD signal from the $m \neq 0$ orbitals must result from the coefficients due to the radial integrals in the actual atoms. These coefficients can be numerically calculated, and typically the $l+1$ channel is the dominant one \cite{Goldberg1981}.

\section{Acknowledgements}

I would like to thank H. Ebert, J. Henk, S. Moser and S. Nemsak for fruitful discussions.

%

\end{document}